\documentclass[pdflatex,sn-mathphys-num]{sn-jnl}

\usepackage{graphicx}%
\usepackage{multirow}%
\usepackage{amsmath,amssymb,amsfonts}%
\usepackage{amsthm}%
\usepackage[scr=rsfs]{mathalpha}
\usepackage[title]{appendix}%
\usepackage{xcolor}%
\usepackage{textcomp}%
\usepackage{manyfoot}%
\usepackage{booktabs}%
\usepackage[ruled,vlined]{algorithm2e}
\usepackage{listings}%

\theoremstyle{thmstyleone}%
\theoremstyle{thmstyletwo}%

\theoremstyle{thmstylethree}%

\begin{document}

\title[CompProv Provenance Graphs]{CompProv Produces Machine Readable Graphs Encoding Microscopic Algebraic Provenance for Reproducible Computation}

\author*[1]{\fnm{Minas} \sur{Abramyan}}\email{minas.abramyan@gmail.com}

\author*[1]{\fnm{Mohammed} \fnm{Alaa} \sur{Ala’anzy}}\email{m.alanzy@ieee.org}

\author[2]{\fnm{Nasir} \sur{Saeed}}\email{mr.nasir.saeed@ieee.org}

\affil*[1]{\orgdiv{Department of Computer Science}, \orgname{SDU University}, \orgaddress{\city{Kaskelen}, \postcode{040900}, \country{Kazakhstan}}}

\affil[2]{\orgdiv{Department of Electrical and Communication Engineering}, \orgname{United Arab Emirates University (UAEU)}, \orgaddress{\city{Al Ain}, \state{Dubai}, \country{UAE}}}

\abstract{
The reliability of computational results in scientific research and financial modeling increasingly depends on verifiable traceability, not merely on trust in a reported output. Existing provenance systems operate at the granularity of files, datasets, or pipeline stages, leaving the internal sequence of algebraic transformations connecting an algorithm's inputs to its outputs unrecorded; a rounding error, an undocumented substitution, or a missing intermediate value can propagate to a final result with no recoverable trace. This work presents CompProv, a Java-based, audit-oriented provenance framework that captures lineage at the atomic level of individual algebraic operations by encapsulating numerical values in high-precision wrapper objects, producing a serializable Calculation Provenance Graph (CPG) that persists as a self-contained artifact rather than a discarded byproduct. The framework is evaluated through three heterogeneous case studies: a decentralized-finance NAV calculation, a reconstruction of an interferometric gauge-block calibration in metrology under explicitly documented input assumptions, and a hydrological model performance evaluation. Deterministic replay reproduced each result exactly in a fresh environment, and CPG-based input substitution supported sensitivity analysis without exposing the underlying source code. These results indicate that, provided the CompProv runtime and wrapper classes are available in the replay environment, a CPG allows numerical integrity to be audited without disclosing proprietary business logic, and that its self-contained structure supports temporal auditability once an execution environment has become deprecated. This framework establishes an operation-level foundation for auditable-by-design computational systems, with scaling to high-throughput computing identified as a direction for future work.}

\keywords{Audit, Data provenance, Metadata, Reproducibility, Transparency}



\maketitle

\section{Introduction}

Reproducibility in computational science and finance depends on the ability 
to verify not just what a computation produced, but precisely how each 
intermediate value was derived. Despite advances in workflow management and 
data lineage systems \cite{ikeda2009data, ahmad2021scientific}, this internal arithmetic record remains largely 
inaccessible. Current provenance frameworks operate at the granularity of 
files, datasets, or pipeline stages~\cite{soiland2022packaging, 
leipzig2021role, maghami2024extensible, ludascher2006scientific}, treating the computational unit 
itself as an opaque process. The inputs and outputs of an algorithm may be 
logged, but the sequence of algebraic transformations that connects them is 
not captured. This gap is consequential: a rounding error, an unauthorized 
coefficient substitution, or an undocumented intermediate approximation can 
propagate silently to a final result with no recoverable trace.

\begin{figure}
    \centering
    \includegraphics[width=0.7\linewidth]{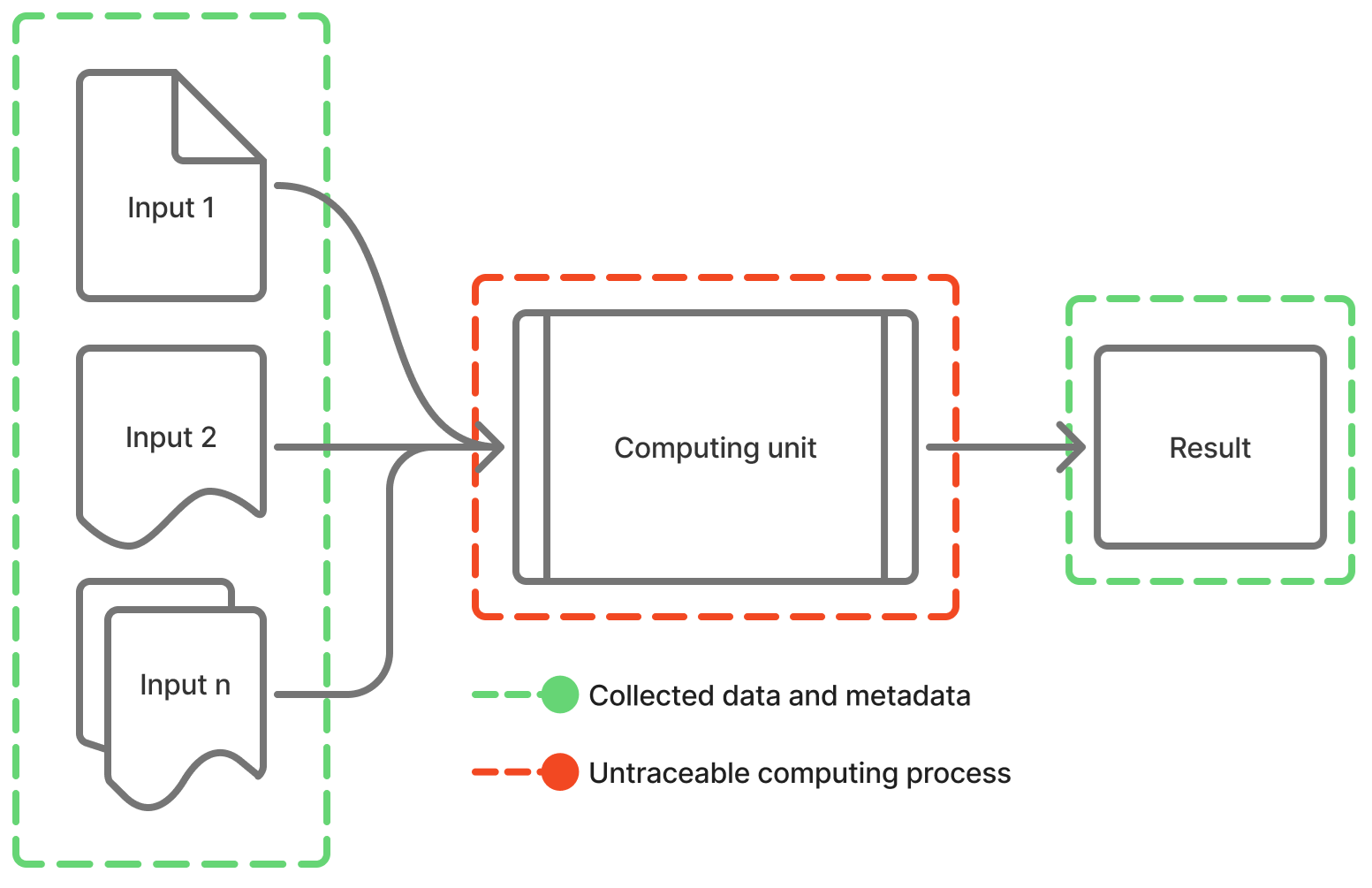}
    \caption{Untraceable computing process: external provenance collectors 
    cannot recover variables that were not explicitly logged during execution.}
    \label{fig:transparency}
\end{figure}

Capturing this level of detail is difficult for a structural reason, not 
merely a practical one. Johnson and Disney~\cite{johnson1999critical} showed 
that manual data recording in software processes introduces systematic 
distortion, what they termed measurement dysfunction, that undermines 
analysis regardless of the recorder's intent; any scheme that depends on a 
developer remembering to log a value inherits this weakness by construction. 
Compiled and interpreted execution compounds the problem: a running program 
computes with intermediate values that exist only in registers and stack 
frames and discards them the moment the next instruction overwrites them, so 
nothing outside the process observes an operation unless something was 
deliberately built to intercept it at the moment it happened. The tools 
capable of that interception, surveyed in Section~\ref{Sec: related}, have 
historically been built for purposes other than auditing, tracing execution 
schedules for debugging, computing gradients for training, or diagnosing 
rounding error, so the interception exists but the resulting record does not 
persist as an audit artifact. Figure~\ref{fig:transparency} illustrates the 
consequence: whatever a provenance collector did not explicitly capture at 
execution time is permanently unrecoverable afterward, no matter how 
completely the surrounding files, environment, or metadata are curated. At a 
larger scale, this same structural gap underlies the reproducibility crisis 
in computational research~\cite{gowda2024symbolic}: when the arithmetic units 
of a computation are not captured, the provenance chain built on top of them 
is unverifiable regardless of how well the layer above it is documented.

Existing systems address this gap at a level of abstraction that stops short 
of the arithmetic. Industrial and scientific lineage systems track 
dependencies between datasets, files, and pipeline stages~\cite{jacques2025unified, 
w3c_prov_dm, ludascher2006scientific, wittner2022lightweight, 
wilkinson2025applying, wagner2022fairly, chirigati2013reprozip, 
villamar2025metadata}, treating each stage as an opaque unit whose internal 
correctness is assumed rather than recorded. A smaller body of work reaches 
inside that unit, either by instrumenting scripts non-intrusively at the 
level of function calls~\cite{murta2015noworkflow, namaki2020vamsa, 
gregori2025llmguided} or by building transient, operation-level computation 
graphs for an entirely different purpose, gradient propagation, in 
automatic-differentiation engines such as PyTorch and 
JAX~\cite{paszke2019pytorch}. A third strand instruments the same 
fine-grained arithmetic to diagnose numerical accuracy rather than to record 
provenance~\cite{jezequel2008cadna, denis2016verificarlo, 
sanchezstern2018finding, eijkhout2009standard, shanmugavelu2024impacts}. 
Section~\ref{Sec: related} situates each of these families in detail and 
shows that none combines operation-level granularity with a persistent, 
replayable, audit-oriented artifact.

The gap that remains, then, is not the absence of fine-grained 
instrumentation, several of the systems above already achieve it for their 
own purposes, but the absence of a mechanism that treats the record of that 
instrumentation as a first-class, persistent output rather than a by-product 
consumed once and discarded. A workflow system's log answers what stage 
produced a file; an automatic-differentiation tape answers what gradient a 
computation implies; neither answers what sequence of algebraic operations, 
applied to which specific input values, produced a specific number, in a 
form that a person with no access to the original source code can still 
open, inspect, and independently replay years after the environment that 
generated it has changed.

In this work the \texttt{CompProv} framework is presented. It is designed for microscopic algebraic provenance; the intrinsic, operation-level capture of numerical lineage 
within a running computation. Numerical values are encapsulated as 
high-precision objects (\texttt{BigDecimal}, \texttt{BigInteger}) that 
automatically record each algebraic transformation they undergo. The result 
is CPG: a serializable, self-contained Directed Acyclic Graph (DAG) encoding every variable, operation, and dependency 
in the computation. Where the computation graph in an AD engine is a 
means to an end, CPG is the end product itself. The CPG can be replayed deterministically, substituted 
with new inputs for sensitivity analysis, and shared with auditors without 
disclosing the proprietary business logic that produced it, provided that 
the CompProv runtime and the corresponding wrapper classes are available in 
the replay environment; what is decoupled from disclosure is the domain 
algorithm, not the provenance infrastructure itself. Both guarantees follow 
directly from where the capture happens: because metadata recording is 
triggered by the act of performing an operation on a wrapped value, rather 
than by a separate call a programmer must remember to make, coverage is a 
structural property of the wrapper type rather than a discipline the 
developer must maintain, and the same immutability that keeps the resulting 
graph acyclic also makes it safe to persist and replay without a later 
mutation invalidating an earlier snapshot.

The specific contributions of this work are:
\begin{enumerate}
    \item A microscopic provenance architecture in which metadata capture 
    is intrinsic to arithmetic execution, not layered externally, closing 
    the gap between operation-level AD graphs and auditable provenance records.
    \item A Governance-by-Design mechanism that transforms numeric primitives 
    into immutable, metadata-carrying objects, eliminating the dependency on 
    developer-managed logging.
    \item A unified, serializable CPG format with a deterministic replay 
    mechanism and input substitution support, enabling independent verification 
    and sensitivity analysis without access to the underlying implementation logic.
    \item Empirical validation across three heterogeneous domains, decentralized finance (NAV calculation), physical metrology 
    (interferometric gauge block calibration), and environmental modeling 
    (hydrological KGE evaluation), demonstrating cross-domain applicability 
    and exact replay fidelity.
\end{enumerate}

The remainder of this paper is organized as follows. Section~\ref{Sec: related} 
reviews existing provenance, lineage, and instrumented-arithmetic approaches 
and identifies the combination of properties that motivates CompProv's 
design. Section~\ref{Sec: methods} 
describes the CompProv architecture, the Calculation Provenance Graph data 
model, and the deterministic replay and input-substitution mechanisms. 
Section~\ref{Sec: results} reports the three case studies and the complexity 
analysis. Section~\ref{Sec: discussion} interprets these findings against 
the existing provenance and instrumented-arithmetic literature and states 
the boundaries of what has been demonstrated. Section~\ref{Sec: threats} 
discusses threats to validity, and Section~\ref{Sec: conclusion} concludes 
the paper.

\section{Related Work}\label{Sec: related}

Every account of computational provenance draws a line at some level of 
granularity below which the record stops and an untraced computation begins. 
That line falls at the level of files and datasets in industrial lineage 
systems, at the level of pipeline stages in scientific workflow tools, and, 
in a smaller body of work, at the level of individual operations inside a 
running program; each depth carries a different notion of what counts as an 
adequate audit trail, a distinction formalized for relational data through 
the why-, how-, and where-provenance semirings of database 
theory~\cite{cheney2009provenance} and surveyed more broadly across 
applications, forms, and system resources by Herschel, Diestelk\"amper, and 
Ben Lahmar~\cite{herschel2017survey}. None of these formalisms, however, was 
designed with imperative numerical computation as its subject: a database 
query has a fixed algebraic structure to instrument, while an arbitrary 
program's arithmetic does not, which is the gap the following three families 
address from different directions.

\subsection{Coarse-Grained Lineage, Workflow, and Environment-Level Provenance}

Industrial-scale lineage systems accept incompleteness as the price of 
scale. Meta's Unified Lineage System~\cite{jacques2025unified} tracks data 
flow across graphs with billions of nodes, recording that a table or 
artifact was produced by a given process, but leaving the internal 
correctness of that process unverified; completeness below the level of the 
dataset is explicitly out of scope, not an oversight. Scientific workflow 
tools inherit a related trade-off from the opposite direction: rather than 
sacrificing depth for scale, they sacrifice depth for portability, wrapping 
computational modules so that files and configuration parameters can be 
tracked without requiring changes to the modules 
themselves~\cite{ludascher2006scientific, wittner2022lightweight, 
ahmad2021scientific}. The W3C PROV data model~\cite{w3c_prov_dm} formalizes 
this same file- and activity-level view as an interchange standard rather 
than a specific system, which is why it recurs across the 
FAIR-computational-workflow literature~\cite{wilkinson2016fair, 
wilkinson2025applying, goble2020fair, welter2023fair, wagner2022fairly, 
soiland2022packaging, maghami2024extensible, di2026code} as a common 
vocabulary for describing what a pipeline did without describing how any 
single stage computed its result.

A complementary strategy avoids modifying the pipeline altogether by 
harvesting whatever the environment already exposes. ReproZip~\cite{chirigati2013reprozip} 
intercepts operating-system calls to capture the files, libraries, and 
configuration a program touches, then bundles the original code together 
with that environment into a package that a reviewer can install and re-run 
elsewhere; the resulting artifact is genuinely self-contained, but the 
self-containment is at the level of the software environment, not the 
arithmetic, so replaying it means re-executing the original program rather 
than replaying a recorded operation sequence. The Archivist 
framework~\cite{villamar2025metadata} takes a similarly non-invasive 
approach for simulation output, harvesting runtime logs and configuration 
state without touching the simulation's source code. Both illustrate the 
same structural limit through different mechanisms: a variable that the 
running program never explicitly wrote to a log, a file, or a monitored 
system call is invisible to the collector no matter how faithfully 
everything else is captured, exactly the gap that leaves seven of the 
thirteen calibration inputs in Section~\ref{Sec: results} unrecoverable from 
the published record and the derivation of the twelve hydrological KGE 
values undocumented in their source study.

Among this family, Archivist~\cite{villamar2025metadata} is both the most 
recent representative and the one adopted here as an empirical baseline, 
since Section~\ref{Sec: results} reconstructs its published hydrological 
benchmark directly rather than citing its limitations in the abstract.

\subsection{Fine-Grained and Operation-Level Provenance}

A much smaller body of work captures provenance below the level of a 
pipeline stage, inside the body of a function or a single script execution, 
and this family divides further by what the resulting record is for. 
noWorkflow~\cite{murta2015noworkflow} instruments Python scripts 
non-intrusively through abstract-syntax-tree analysis and runtime profiling, 
recording function activations, arguments, and return values without 
requiring the scientist to annotate or restructure the code; because it 
targets user-defined functions rather than every arithmetic operation inside 
them, a single call that performs thousands of floating-point operations is 
recorded as one node, and the granularity is set by how the programmer 
happened to factor the script into functions rather than by the arithmetic 
itself. Vamsa~\cite{namaki2020vamsa} narrows the same script-level approach 
to a specific question, statically analyzing data-science scripts to 
determine which columns of a source dataset were used to derive a trained 
model's features and labels, which is enough to answer fairness and 
compliance questions about column-level lineage without ever executing the 
script, but stops well short of recording the operations performed on the 
values within those columns. A more recent system, 
PROLIT~\cite{gregori2025llmguided}, narrows the granularity gap 
deliberately by letting a user choose among four explicit levels, from a 
coarse sketch down to a full per-cell dependency graph over pandas 
dataframes, using a large language model to segment and annotate the 
pipeline before execution; at its finest level it approaches the atomicity 
CompProv targets, but the resulting record is a queryable graph-database 
representation produced by an LLM-assisted, code-rewriting front end rather 
than a machine-executable replay artifact, and reproducing an entire run 
from it is not among its stated goals.

None of these three systems is a serialized, replayable execution record in 
the sense Section~\ref{Sec: methods} defines, which raises a terminological 
point worth making explicit given how one term recurs across two largely 
separate literatures. A substantial body of systems research addresses 
\emph{deterministic replay} as the problem of reproducing a non-deterministic 
multi-threaded or distributed execution exactly, typically by recording 
thread interleavings, memory accesses, or system-call orderings so that a 
debugger can step through the same buggy run twice~\cite{chen2015deterministic}. 
CompProv's replay mechanism (Algorithm~\ref{alg:replay-mechanism}) addresses 
a different problem: the recorded Snapshot already fixes every operation and 
its inputs, so no scheduling non-determinism remains to reproduce, and 
replay instead means re-resolving each recorded operation against the 
wrapper classes available in the new environment. The two notions of replay 
share a name and a general goal, faithfully reproducing a past execution, 
but operate on different objects and solve different problems, and neither 
literature should be read as a competing solution to the other's question.

Tape-based automatic-differentiation engines such as PyTorch and 
JAX~\cite{paszke2019pytorch} remain the closest architectural relative to 
CompProv within this family, building genuinely operation-level graphs 
through the same operator-overloading mechanism CompProv uses, and are 
retained as the family's representative baseline in Table~\ref{tab:comparison} 
for that reason. PROLIT is the most recent system in the broader 
fine-grained family, but its target language, runtime model, and goal, 
explaining a data-preparation pipeline's effect on tabular data rather than 
persisting a replayable numerical computation, differ enough from 
CompProv's that adopting it as a head-to-head performance baseline would 
compare two systems built for different jobs rather than two solutions to 
the same one.

\subsection{Numerical Accuracy and Rounding-Error Diagnostics}

A separate strand of work instruments arithmetic not to record what happened 
but to diagnose how much the answer can be trusted. Discrete stochastic 
arithmetic tools such as CADNA estimate round-off error propagation by 
running perturbed variants of a computation and comparing the significant 
digits that survive across them~\cite{jezequel2008cadna}, and Verificarlo 
applies a comparable Monte Carlo Arithmetic strategy at the compiler level 
to quantify the sensitivity of C, C++, and Fortran codes to 
rounding~\cite{denis2016verificarlo}. Herbgrind takes a complementary 
dynamic-analysis approach, tracing the propagation of floating-point error 
through a program to localize the specific operation responsible for a loss 
of accuracy~\cite{sanchezstern2018finding}, and Eijkhout and Fuentes propose 
a standard for attaching this kind of numerical metadata to values in the 
first place~\cite{eijkhout2009standard}. The practical stakes of this 
diagnostic question have only grown with scale: a recent large-scale study 
of floating-point non-associativity across HPC and deep-learning workloads 
shows that summation order alone can shift results enough to threaten 
reproducibility on modern accelerated hardware~\cite{shanmugavelu2024impacts}, 
exactly the kind of silent divergence a CPG cannot by itself explain, since 
CompProv records which operations executed in which order, not whether a 
different, equally valid order would have changed the outcome.

This family answers a different question from the one CompProv poses. Its 
tools ask whether a result is numerically trustworthy; CompProv asks whether 
a result's derivation is recorded and auditable at all, a question that 
remains open even when every operation executes with full precision and no 
rounding error occurs. The two questions are complementary rather than 
competing, and Verificarlo, as the most actively maintained and most recent 
widely used system in this family, is not adopted as a baseline here for 
that reason: a benchmark comparing CompProv's tracing overhead against 
Verificarlo's Monte Carlo Arithmetic overhead would compare the cost of two 
different instruments built to measure two different things.

\begin{table}[htbp]
\centering
\footnotesize
\caption{Comparison of CompProv with representative provenance, lineage, and 
computation-graph systems}
\label{tab:comparison}
\begin{tabular}{lcccccc}
\hline
\textbf{System} & \textbf{Granu-} & \textbf{Serial-} & \textbf{Replay} & 
\textbf{Audit} & \textbf{Source-} & \textbf{Overhead} \\
 & \textbf{larity} & \textbf{izable} & & \textbf{metadata} & 
\textbf{code-free} & \textbf{reported} \\
\hline
Meta ULS~\cite{jacques2025unified} & Dataset & No & No & Partial & No & None \\
W3C PROV~\cite{w3c_prov_dm} & File & Yes & No & Contextual & No & N/A \\
ReproZip~\cite{chirigati2013reprozip} & Environment & Yes & Partial\tnote{a} & Contextual & No & None \\
Archivist~\cite{villamar2025metadata} & File/Log & Partial & No & Contextual & No & None \\
noWorkflow~\cite{murta2015noworkflow} & Function & Yes & No & Contextual & No & Non-burd.\tnote{b} \\
PyTorch AD~\cite{paszke2019pytorch} & Operation & No & No & None & Yes & N/A\tnote{c} \\
\textbf{CompProv} & \textbf{Operation} & \textbf{Yes} & \textbf{Yes} & 
\textbf{Full} & \textbf{Yes} & \textbf{2.1--10.4$\times$\tnote{d}} \\
\hline
\end{tabular}
\begin{tablenotes}
\small
\item[a] Re-executes the original packaged program in a reconstructed 
environment rather than replaying a recorded operation-level trace.
\item[b] Self-reported by the system's authors without a quantified figure.
\item[c] Transient tape, not benchmarked as a provenance-tracing overhead.
\item[d] Quantified in Section~\ref{Subsec: complexity}; range reflects 
unfolded-to-folded tracing across two benchmark payloads of differing 
computational intensity. ``None'' indicates the source publication reports 
no overhead figure of any kind, quantified or otherwise.
\end{tablenotes}
\end{table}

Read together, these three families cover five properties that an 
operation-level, audit-oriented provenance mechanism would need: 
operation-level granularity, a persistent serializable artifact, 
deterministic replay of that artifact, semantic audit metadata attached to 
each value, and decoupling from the original proprietary source; but no 
single system in Table~\ref{tab:comparison} provides more than three of the 
five at once. Coarse-grained lineage and workflow systems provide a 
serializable, audit-friendly artifact but stop at the file or pipeline 
boundary; automatic-differentiation engines reach operation-level 
granularity but discard the graph as a transient computational device; 
script-level tools such as noWorkflow, Vamsa, and PROLIT narrow that gap for 
specific languages or specific questions without producing a self-contained, 
replayable record; and numerical-diagnostic tools instrument the same 
arithmetic CompProv targets but answer a question about accuracy rather than 
about auditability. CompProv is built to close exactly this combination, not 
by improving any one existing mechanism, but by making metadata capture and 
persistence intrinsic to the arithmetic itself, a design choice whose 
concrete architecture, data model, and replay algorithm are developed next.

\section{CompProv Architecture: Microscopic Provenance Capture and Deterministic Replay}\label{Sec: methods}
\begin{figure}[htbp]
    \centering
    \includegraphics[width=0.8\linewidth]{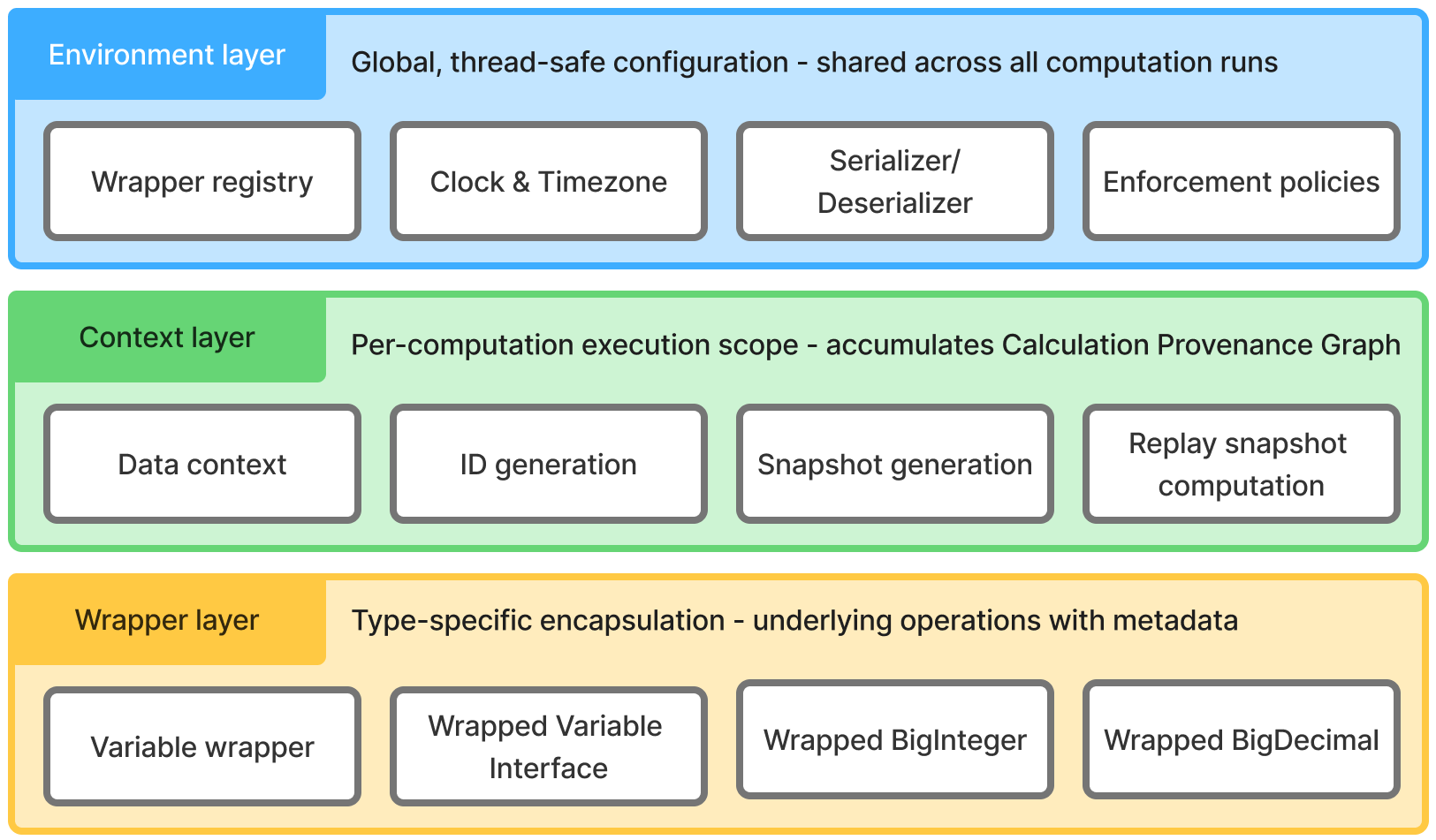}
    \caption{CompProv Framework - layered architecture}
    \label{fig:architecture}
\end{figure}

The proposed \texttt{CompProv} framework instruments numerical computations by making the metadata collection process an intrinsic part of the computational process. Figure \ref{fig:architecture} illustrates the architecture organized into cooperative layers.
The layers are described as follows:
\begin{enumerate}
    \item The environment Layer establishes a global registry of data types and enforcement policies, ensuring consistency across disparate computational nodes and providing shared configuration for reproducible calculation.
    \item The context layer serves as a temporal execution container, accumulating the CPG as execution proceeds. Because Variable and Operation nodes are never mutated once created, concurrent threads appending new nodes cannot invalidate one another's view of previously recorded state, which allows the layer to manage concurrent operations without a global lock. A 4-threaded execution benchmark (Section~\ref{Subsec: complexity}) shows that wrapped execution scales in line with native execution under this workload, which is consistent with this design rationale; that benchmark measures throughput under one concurrency pattern, however, and does not by itself constitute formal verification of thread safety or reentrancy across arbitrary usage patterns, which remains a direction for future work.
\end{enumerate}

Initially, the programmer interacts with the framework through the environment. The environment layer is used as an entry point to create or derive the context layer. The most important function context layer provides is a wrap function. The wrap function identifies the corresponding wrapper depending on the input value. In the environment’s registry and puts the wrapped variable with descriptors and other details into context. The programmer then works with the wrapped variable. The wrapped variable provides value-specific functions to interact with the underlying value and implements the wrapped variable interface. The wrapped variable interface has \texttt{getFunction} function which returns the implementation of the underlying function by the operation descriptor. The \texttt{getFunction} function is crucial for the reproducibility of the calculation.

\begin{figure}[htbp]
    \centering
    \includegraphics[width=0.8\linewidth]{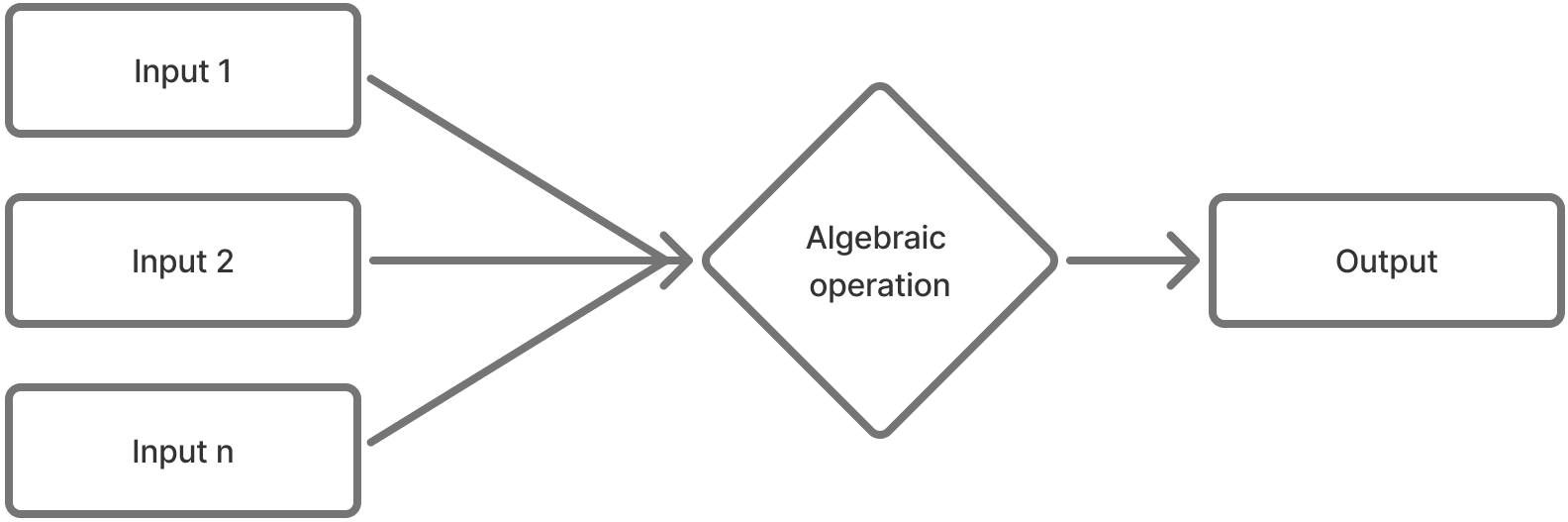}
    \caption{CPG fragment}
    \label{fig:graph-fragment}
\end{figure}

The CPG is formalized as a DAG with two node types: a Variable node represents an immutable variable state at a moment in time and an Operation node is also immutable, it is the specific action executed at the moment. The directed edges connect the variables and operations that form the calculation flow. Variables and operations immutability guarantees acyclicity of the graph. Figure \ref{fig:graph-fragment} represents a fragment of the CPG.

The graph is encoded implicitly through Operation Records, constituting a functional mapping:
\begin{equation}\label{Eq: functional-mapping}
   f:(I_{1}, I_{2},...,I_{n})\rightarrow O
\end{equation}
where \textit{I} represent the identifiers of the input variables and \textit{O} represents the resulting output. 

Every variable in the CPG is augmented with an immutable Variable Track object. This object is designed to provide human- and machine-readable metadata essential for CPG calculation replay. It offers comprehensive context for every value, such as its origin, name, and type, while allowing developers to attach any additional information required for a transparent and robust audit trail. Table \ref{tab:variable-track} lists all fields of the variable track. 
\begin{table}[htb]
\centering
\caption{Fields of variable track}
\label{tab:variable-track}
\begin{tabular} { | p{1.4cm} | p{2cm} | p{8cm}| } \hline
\textbf{Field} & \textbf{Type} & \textbf{Description} \\ \hline
id & String & Prefixed string ID, format: i\_N or o\_N where i/o points to kind: INPUT/OUTPUT and N is numericId. \\ \hline
numericId & Int & Autogenerated sequential integer, source of truth for ordering. \\ \hline
createdAt & ZonedDateTime & High-resolution creation timestamp (UTC). \\ \hline
kind & Enumeration of INPUT or OUTPUT & This specifies the variable type: INPUT denotes a value defined as an external parameter or constant, whereas OUTPUT is generated as the result of a computational operation. \\ \hline
descriptor & Descriptor & A Descriptor contains a human-readable name and a Meta object. The latter serves as a thin wrapper for an ordered key-value map, enabling the attachment of additional context and granular metadata. \\  \hline
valueClass & String & The fully-qualified Java class of the wrapped value plays a crucial role in CPG replay; facilitates the dynamic binding of the variable to its native computational type during replay and dictates exactly how the variable should be represented programmatically. \\ \hline
\end{tabular}
\end{table}

Every recorded algebraic operation is defined by an Operation Track object. Beyond linking inputs to their respective results, this object encapsulates the human and machine-readable metadata essential for automated CPG replay and provides the necessary context for an auditor to understand the transformation of input variables into the final output. The complete set of operation track fields is detailed in Table \ref{tab:operation-track}.

\begin{table}[htb]
\centering
\caption{Fields of operation track}
\label{tab:operation-track}
\begin{tabular} { |p{1.6cm} | p{0.15\linewidth} | p{8cm} | } 
\hline
\textbf{Field} & \textbf{Type} & \textbf{Description} \\ \hline
id & String & Prefixed string ID, format: op\_N where N is numericId. \\ \hline
numericId & Int & Autogenerated sequential integer, source of truth for ordering. \\ \hline
startedAt & ZonedDateTime & High-resolution timestamp (UTC) of operation starting. \\ \hline
finishedAt & ZonedDateTime & High-resolution timestamp (UTC) of operation finishing. \\ \hline
descriptor & Descriptor & The operation descriptor includes the operation name and metadata, which are designed to provide a human-readable description of the computational transformation; it serves as the identifier for the functional interface implementation during CPG replay. \\ \hline
wrapperClass & String & The fully-qualified Java class of the wrapper. \\ \hline
\end{tabular}
\end{table}

Ultimately, the CPG is represented as a Snapshot, a serializable, self-contained manifestation of the entire provenance graph. The Snapshot comprises three primary fields:
\begin{enumerate}
    \item descriptor - contains a name and metadata designed to provide human-readable context.
    \item variables - an ordered list of variable records. Each record pairs a Variable Track with its concrete Java value object (such as a BigDecimal or a custom domain object). These are sorted by their numericId to ensure a consistent reference structure.
    \item operations - an ordered list of operation records. Each pairs an Operation Track with an ordered map of arguments (mapping argument names to variable IDs) and a resultId string (points to result variable). These are sorted by numericId, which corresponds to their topological order since operations are appended chronologically during execution.
\end{enumerate}

The self-contained guarantee of the Snapshot depends on the operational scope encapsulated within the graph. When a computation relies exclusively on the core, built-in algebraic operations and environment types natively provided by the framework, the Snapshot is entirely self-contained, since the baseline infrastructure already possesses every operation descriptor and lookup mechanism required to deterministically replay it without external dependencies. When the CPG instead encapsulates custom user-defined operations or specialized wrapper classes, as in the multi-domain case studies in Section~\ref{Sec: results}, this guarantee narrows in the precise way described next.

The Snapshot can be serialized as a JSON \cite{bray2014javascript} object (Listing \ref{lst:json-snapshot}) to be stored and utilized for calculation provenance and reproducibility purposes
\begin{minipage}{\linewidth}
\footnotesize
\begin{lstlisting}[caption={JSON snapshot serialization example}, label={lst:json-snapshot}] 
  "descriptor" : { "name" : "NAV", "meta" : { "id" : "b09e8391e9f9 " } },
  "variables" : [ { 
...
    "track" : { "id" : "i_2", "numericId" : 2, 
        "createdAt" : "2026-03-14T02:09:32.1850532Z", 
        "kind" : "INPUT", 
      "descriptor" : { "name" : "BTC price", "meta" : {"origin" : "Binance"} },
      "valueClass" : "java.math.BigDecimal"
    },"value" : "68792.45" },
...
  ], 
  "operations" : [ {
...
    "track" : { "id" : "op_2", "numericId" : 2, 
        "startedAt" : "2026-03-14T02:09:32.1850532Z", 
        "finishedAt" : "2026-03-14T02:09:32.1850532Z",
      "descriptor" : { "name" : "add", 
        "meta" : { "formula" : "(a+b)mc" } },
      "wrapperClass" : "io.compprov.core.wrappers.WrappedBigDecimal"
    },"arguments" : [
      {"key":"a", "value":"i_2"}, 
      {"key":"b", "value":"o_4"}, 
      {"key": "mc", "value":"i_1" }
    ], "resultId" : "o_5" } 
...
  ] }
\end{lstlisting}
\end{minipage}

The computational reproducibility of the serialized CPG is guaranteed by the persistence of the variable states and the chronological sequence of the operation records. Algorithm \ref{alg:replay-mechanism} represents the replay mechanism.

A precise statement of what "self-contained" means for the Snapshot is necessary before presenting the replay mechanism, since Algorithm~\ref{alg:replay-mechanism} depends on components that are not shipped inside the Snapshot itself. The Snapshot is self-contained with respect to the \emph{data} of the computation: every variable value, its complete Variable Track metadata, and the full, chronologically ordered sequence of Operation Track records are embedded in the serialized artifact, so no external file, database, or access to the original business logic is required to interpret what happened. Replay, however, is not independent of the CompProv \emph{infrastructure} whenever custom or extended components are involved: Step~2 of Algorithm~\ref{alg:replay-mechanism} resolves each variable's \texttt{valueClass} against the Environment registry to locate the matching wrapper class, and Step~3.4 of the \texttt{ExecuteOperation} procedure resolves each operation's descriptor against the calling object's \texttt{LookupFunction} table to locate the corresponding operation implementation. Both the wrapper class and the operation implementation must therefore be present in the replay environment whenever the recorded computation used custom domain-specific types or extended operations, as in the case studies of Section~\ref{Sec: results}. What deterministic replay eliminates is the dependency on the original proprietary business logic that orchestrated these calls, not the dependency on the CompProv runtime itself. In this respect, deterministic replay refers specifically to the exact reproduction of computed values together with the preservation of the recorded variable and operation metadata; it does not imply that a non-deterministic external value, such as a market price fetched at execution time, is re-queried or re-verified against its original source during replay, since such a value is itself captured as an immutable input at the moment it was first wrapped.

\begin{algorithm}[htbp]
\caption{The replay mechanism}
\label{alg:replay-mechanism}
\footnotesize
\SetKwFunction{FReplay}{ReplayCPG}
\SetKwFunction{FExecute}{ExecuteOperation}
\SetKwProg{Fn}{PROCEDURE}{:}{END PROCEDURE}

\Fn{\FReplay{Snapshot}}{
    \tcp{Step 1: New computation context is created}
    $Context \leftarrow \textbf{new } \text{ComputationContext()}$\;
    \BlankLine
    \tcp{Step 2: Every variable is re-wrapped into the context; the wrapper class must be registered in the replay environment}
    \ForEach{Var \textbf{in} Snapshot.Variables}{
        $WrapperClass \leftarrow \text{Environment.Registry.Find}(Var.valueClass)$\;
        $WrappedVar \leftarrow \textbf{new } WrapperClass(\\
            Var.id, \\
            Var.value, \\
            Var.metadata)$\;
        $Context.Add(WrappedVar)$\;
    } 
    
    \BlankLine
    \tcp{Step 3: Every operation is re-executed in chronological order}
    \ForEach{Op \textbf{in} Snapshot.Operations}{
        \FExecute{Op, Context}\;
    }
}

\BlankLine 

\Fn{\FExecute{Op, Context}}{
    \tcp{3.1 \& 3.2: Argument IDs and input variables are mapped}
    $Inputs \leftarrow []$\;
    \ForEach{ArgId \textbf{in} Op.ArgumentIds}{
        $Inputs.Append(Context.GetById(ArgId))$\;
    }
    
    \BlankLine
    \tcp{3.3: The first argument is used as a calling object}
    $CallingObject \leftarrow Inputs[0]$\;
    
    \BlankLine
    \tcp{3.4: Locate the operation implementation registered for this operation descriptor}
    $OpImplementation \leftarrow CallingObject.LookupFunction(Op.Descriptor)$\;
    
    \BlankLine
    \tcp{3.5: Operation implementation is executed}
    $ResultValue \leftarrow OpImplementation.Execute(Inputs)$\;
    
    \BlankLine
    \tcp{3.6 \& 3.7: Update states}
    $Op.ActualTimestamp \leftarrow \text{CurrentTimestamp()}$\;
    $ResultVar \leftarrow Context.GetById(Op.ResultId)$\;
    $ResultVar.Value \leftarrow ResultValue$\;
    $ResultVar.ActualTimestamp \leftarrow \text{CurrentTimestamp()}$\;
}
\end{algorithm}

\begin{figure}
    \centering
    \includegraphics[width=1\linewidth]{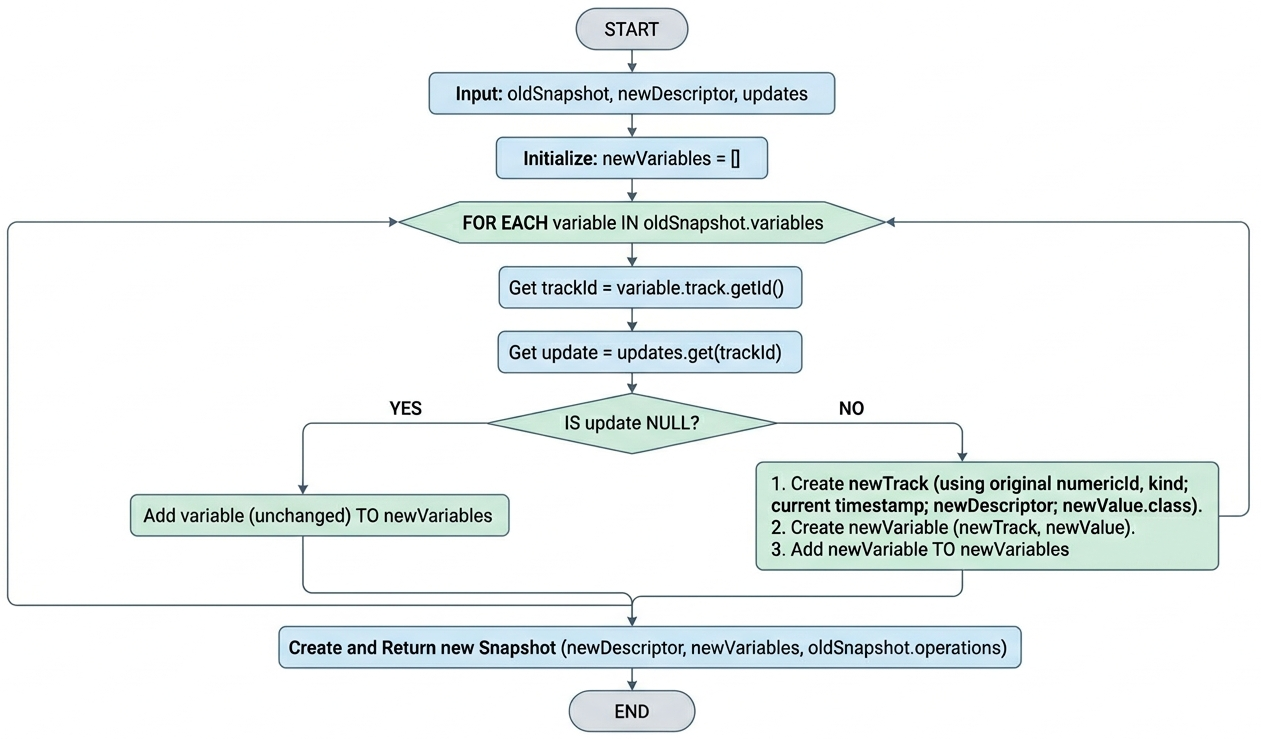}
    \caption{Copy the snapshot with data substitution algorithm}
    \label{fig:copy-with-substitution}
\end{figure}

The framework also supports simulations without source-code interventions. The environment provides a function to substitute input variables. Substitution is done through snapshot duplication; details of the algorithm are reflected in Algorithm \ref{alg:copy-with-substitution} and Figure \ref{fig:copy-with-substitution}. Each entry in the \texttt{updates} map pairs a target variable's track identifier with an \texttt{update} record carrying two fields: \texttt{update.value}, the replacement value itself, and \texttt{update.descriptor}, a caller-supplied label for that specific substitution (for example, "stress-test override"). This keeps the per-variable descriptor decoupled from both the variable's original descriptor and from \texttt{newDescriptor}, which labels the resulting Snapshot as a whole; the three serve distinct, non-overlapping purposes and are never conflated.

\begin{algorithm}[htbp]
\caption{Copy the snapshot with data substitution algorithm}
\label{alg:copy-with-substitution}

\SetKwFunction{FCPY}{copyWith}
\SetKwProg{Fn}{Function}{:}{}
\SetKw{Create}{Create}

\Fn{\FCPY{oldSnapshot, newDescriptor, updates}}{
    \tcp{Initialize result container}
    $newVariables \gets \text{Empty List}$\;
    \BlankLine
    \tcp{Process Variable Substitution}
    \ForEach{$variable$ \textbf{in} $oldSnapshot.variables$}{
        $trackId \gets variable.track.getId()$\;
        $update \gets updates.get(trackId)$\;
        
        \eIf{$update$ \textbf{is} \text{NULL}}{
            \tcp{Retain original variable if no substitution is requested}
            $newVariables.add(variable)$\;
        }{
            \tcp{Only INPUT-kind variables are eligible for substitution}
            \eIf{$variable.track.kind \neq \text{INPUT}$}{
                \Return \text{Error}(\text{"Cannot substitute a computed OUTPUT variable"})\;
            }{
            \tcp{Instantiate a new track that preserves the original id and numericId}
            $newTrack \gets$ \Create New Track(
                \Indp
                id: $variable.track.id$, \\
                numericId: $variable.track.numericId$, \\
                timestamp: $System.currentTime()$, \\
                kind: $variable.track.kind$, \\
                descriptor: $update.descriptor$, \\
                valueClass: $update.value.getClass()$
                \Indm
            )\;
            \BlankLine
            \tcp{Generate the substituted variable object}
            $newVariable \gets$ \Create New Variable(\\
                $newTrack, \\
                update.value$
            )\;
            $newVariables.add(newVariable)$\;
            }
        }
    }
    \BlankLine
    \tcp{Construct the Synthetic Snapshot; newDescriptor labels the substituted scenario as a whole and is distinct from any individual variable's own descriptor}
    \tcp{Operations are preserved to enable Deterministic Replay}
    \Return \Create New Snapshot(\\
        \Indp
        descriptor: $newDescriptor$, \\
        variables: $newVariables$, \\
        operations: $oldSnapshot.operations$
        \Indm
    )\;
}

\end{algorithm}

The framework is implemented in Java (version 17+). Modern Java programming features, such as static imports and object-oriented paradigms, align the CompProv framework with established development practices. These features facilitate a developer-friendly environment, ensuring that the code remains both readable and intuitive to construct. Listing \ref{lst:code-wrapping} demonstrates a code fragment with variables wrapping and algebraic operation execution.

\begin{minipage}{\linewidth}
\begin{lstlisting}[caption={CompProv wrapping variable and operation execution}, label={lst:code-wrapping}] 
    final var lambdaVac3 = ctx.wrapBigDecimal(
        data.fetchLambdaVac3(),
        descriptor(
            "lambda_vac3: Spectra 117A-1 HeNe vacuum wavelength, nm", 
            of("report", "OFS-2024-0002", "date", "2024-05-08")));
    
    final var lambdaVac3um = lambdaVac3.divide(
        cNmToUm, 
        mc, 
        descriptor("lambda_vac3 in um"));
\end{lstlisting}
\end{minipage}

The CompProv framework is distributed as a Maven-managed artifact via Maven Central, enabling seamless integration into standard Continuous Integration/Continuous Delivery (CI/CD) pipelines and Java-based research ecosystems without the need for additional repository configurations. To make the reported experiments independently reproducible, the exact code-and-data snapshot used for the three case studies and the complexity analysis in Section~\ref{Sec: results} is archived separately from the framework's rolling development history and is available online in our repository \footnote{https://github.com/compprov/compprov-core/releases/tag/v0.4.2}.

In addition to the core framework, a dedicated rendering engine has been developed to reconstruct and visualize the CPG from JSON snapshots \footnote{https://github.com/compprov/compprov-render/releases/tag/v0.2.0}. The engine is implemented as a standalone Hypertext Markup Language (HTML) page supported by a suite of JavaScript and Cascading Style Sheets (CSS) files. To ensure accessibility, the interface can be opened directly in any modern web browser without requiring a specialized server environment. A user-friendly interface allows researchers to import CPG data in JSON format and interactively explore the metadata for any variable or operation. Furthermore, the displayed fields are fully customizable and do not require programming expertise to configure (Figure \ref{fig:navgraph}). 

\section{Results}\label{Sec: results}

To evaluate the functional utility and performance of the CompProv framework, three distinct case studies are presented in the domains of decentralized finance, physical metrology, and environmental science. These use cases were selected to demonstrate the versatility of the framework in handling varying data structures, ranging from simple scalar formula chains to complex high-frequency transactional data.

In each evaluation, the main objective is to move beyond the traditional black-box computation model. Instead, the computation context is utilized to wrap primary inputs and intermediate operations, thereby generating a persistent CPG that exists alongside the numerical result. Through this approach, it is ensured that every output is accompanied by a machine-readable audit trail.

\subsection{Provenance in NAV Calculation}

To demonstrate the practical application of the CompProv framework, a simulation of a multi-asset crypto-portfolio NAV calculation was implemented. In this scenario, the system computes the total value in US Dollars (USD) of a portfolio containing various assets held on different platforms such as Binance, stake contracts and the Morpho protocol. The input values are provided in Table \ref{tab:navdata}

\begin{table}[htb]
\centering
\caption{Portfolio Input Variables and Source Metadata}
\label{tab:navdata}
\begin{tabular}{|c|c|c|c|}
\hline
\textbf{Variable} & \textbf{Value} & \textbf{Source Identifier} & \textbf{Metadata Origin} \\ \hline
Bitcoin (BTC) amount  & 2.13      & \texttt{binanceBtcAmount}  & Binance \\ \hline
Ethereum (ETH) amount & 23.34     & \texttt{binanceEthAmount}  & Binance \\ \hline
USD Coin (USDC) amount & 532.9     & \texttt{binanceUsdcAmount} & Binance \\ \hline
Staked ETH amount    & 5.91      & \texttt{stakedEthAmount}  & Stake   \\ \hline
USDC amount          & 221114.9  & \texttt{morphoUsdcAmount}  & Morpho  \\ \hline
BTC/USD rate  & 68989.72  & \texttt{btcUsdPrice}   & Binance  \\ \hline
ETH/USD rate & 2083.31  & \texttt{ethUsdPrice}   & Binance  \\ \hline
USDC/USD rate & 1.01  & \texttt{usdcUsdPrice}   & Binance  \\ \hline
\end{tabular}
\end{table}

A non-invasive wrapping pattern is followed in the implementation, where existing business-domain models Amount, Rate, and Currency are integrated into the framework without the modification of their original source code. This is achieved through a three-step integration:
\begin{enumerate}
    \item Domain types are encapsulated in tracked containers implementing the wrapped variable interface.
    \item Lightweight factories are implemented to instantiate wrapped classes from raw values.
    \item These factories are registered within the environment layer, which serves as a configuration.
\end{enumerate}

This encapsulation strategy serves as a viable compromise for commercial applications where intellectual property must be protected. By decoupling domain models such as \texttt{Amount}, \texttt{Rate}, and \texttt{Currency} from the underlying proprietary algorithms, an organization can maintain the privacy of its internal logic while still providing a complete provenance record. Consequently, the resulting CPG can be replayed and simulated independently of the commercial source code. This enables external auditing and enhances user transparency in high-stakes computations, such as the determination of Total Value Locked (TVL), without exposing sensitive trade secrets. Furthermore, stored CPG addresses the significant challenge of temporal auditability; in high-stakes sectors, performing historical audits often becomes technically unfeasible as original source codes, specific library versions, and execution environments evolve or become deprecated. Because CPG serves as a self-contained, immutable record of the operational logic at the moment of execution, it eliminates the necessity of reconstructing legacy software states. This ensures that the numerical integrity of historical data remains verifiable even years after the original computational environment has been decommissioned.

\begin{figure}
    \centering
    \includegraphics[width=1\linewidth]{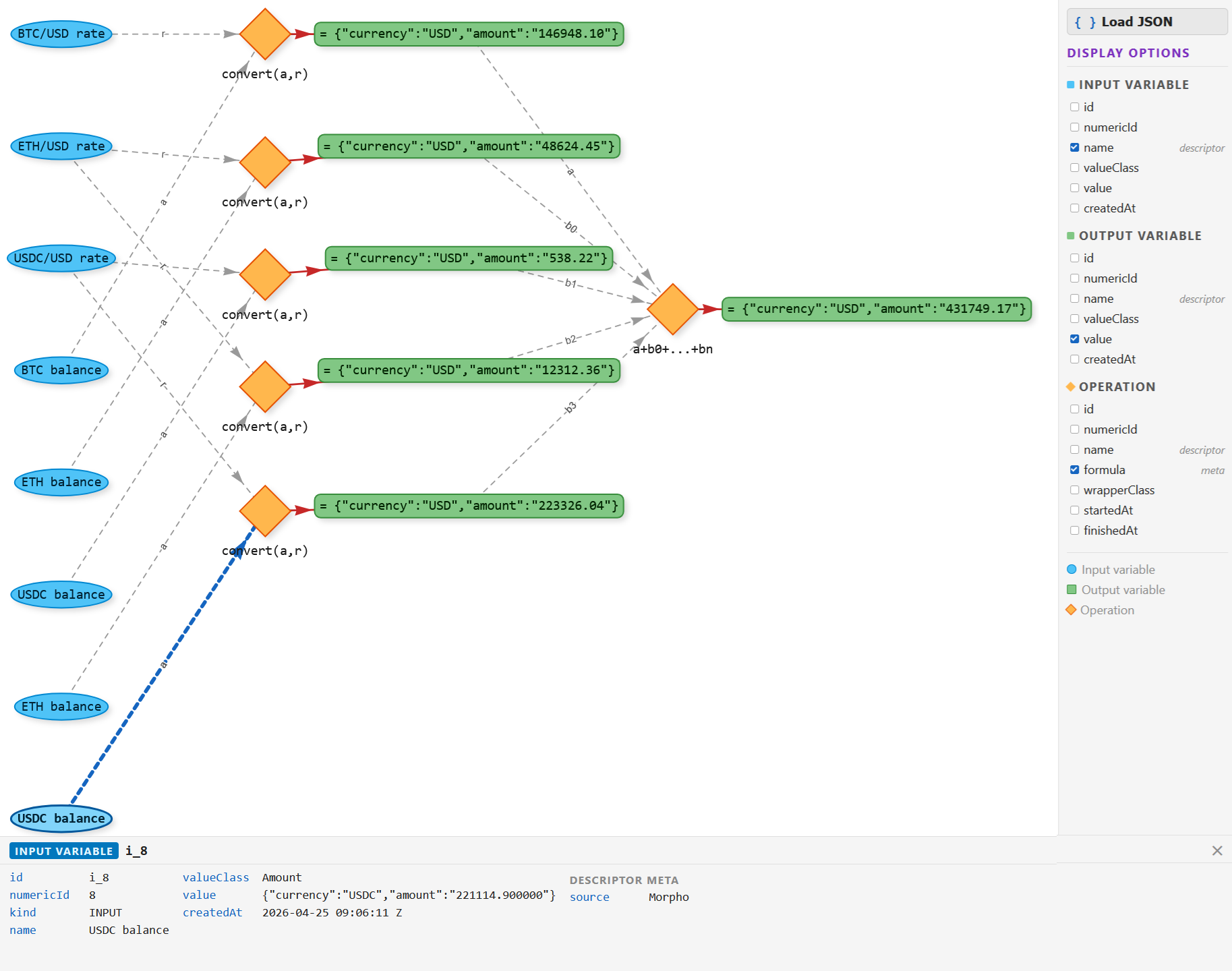}
    \caption{NAV CPG visualization through the rendering engine}
    \label{fig:navgraph}
\end{figure}

In the simulation, atomic-level traceability is achieved through the automatic recording of every algebraic transformation, initiated at the point where external data such as exchange rates are ingested and encapsulated into the framework. Although the retrieval process itself remains external, the resulting values are immediately complemented with comprehensive metadata, including details such as data origin, arrival timestamps, etc. For example, once a Binance-sourced BTC/USD rate is retrieved and enclosed within a tracked container, any subsequent execution of an operation (e.g., the \texttt{Amount\#convert} method) triggers the recording of the execution context and operation details in the CPG. This mechanism ensures that high-resolution timestamps and lineage links are captured for the start and end of the conversion, establishing a persistent operation track that connects the initial balance and the external rate to the final USD result (a visualization of the resulting CPG is provided in Figure \ref{fig:navgraph}).

The integrity of the results was verified using the deterministic replay mechanism. By serializing the CPG into a snapshot, the total computational state was reconstructed in a fresh environment, confirming that the final NAV of 431,749.17 USD remained consistent with the provided inputs and operational logic. This process facilitates a robust Chain-of-Custody Verification, allowing auditors or regulators to re-execute operations and confirm that no unauthorized changes to coefficients or rounding errors occurred during the calculation.

The calculation of the final NAV proceeds through a sequence of operations (\texttt{convert}, \texttt{addBulk}), each of which generates a unique node in the resulting CPG.
Within this structure, atomic lineage is preserved by recording the conversion of BTC to USD not as a simple scalar multiplication, but as a formal functional mapping:
\begin{equation}
    f:(I_{BTC}, I_{rate})\rightarrow O_{USD}
\end{equation}

\begin{figure}
    \centering
    \includegraphics[width=0.7\linewidth]{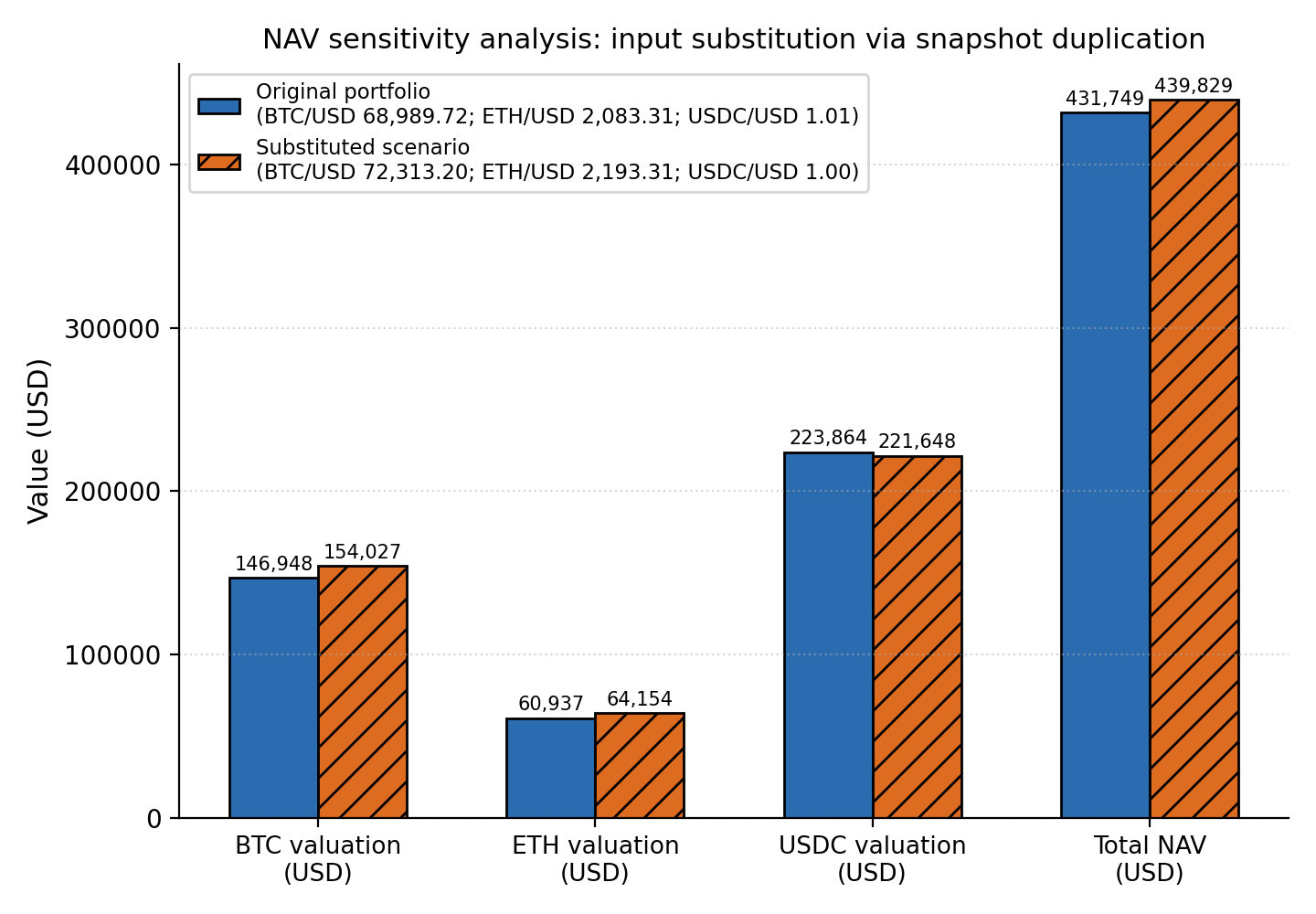}
    \caption{NAV sensitivity analysis}
    \label{fig:sensitivity-analysis}
\end{figure}

Furthermore, semantic preservation is achieved as every intermediate value, such as the USD value of staked ETH, retains its physical nature through encapsulation, thereby preventing the loss of numerical context. By leveraging the CPG snapshot, the framework also enables sensitivity analysis, where specific input variables are substituted to propagate changes through the full graph, allowing for experimentation without the modification of the original source code. Figure~\ref{fig:sensitivity-analysis} compares the original portfolio against a substituted scenario in which only the three exchange rates (BTC/USD, ETH/USD, and USDC/USD) are changed, holding all asset amounts fixed, and reports the resulting per-asset valuations and total NAV for both scenarios.

\subsection{Interferometric Gauge Block Calibration}

The second evaluation applies the CompProv architecture to the strict regulatory environment of physical metrology, through an independent reconstruction of the interferometric gauge block calibration procedure reported in \cite{white2025provenance}. In metrology, an unbroken chain of comparisons is the foundational requirement for measurement uncertainty calculations and international standards compliance \cite{jcgm_vim, jcgm_gum}.

The algebraic workflow of the calibration of The National Research Council of Canada (NRC) into the CompProv framework to measure the 7 mm tungsten carbide gauge block. The measurement relies on the method of exact fractions, comparing three calibrated laser vacuum wavelengths including a primary laser calibrated vacuum wavelength against the artifact to resolve the integer fringe order.

Unlike the financial use case, which required custom business-domain wrappers, this evaluation demonstrates that the framework’s native arithmetic is sufficient to encapsulate highly complex, pure-scalar formula chains. All input values for the gauge block calibration are provided in Table \ref{tab:metrology-inputs}

\begin{table}[htb]
\centering
\caption{Input values for the gauge block calibration}
\label{tab:metrology-inputs}
\begin{tabular}{ | p{0.4\linewidth} | p{0.2\linewidth} | p{0.4\linewidth} | } \hline
\textbf{Description} & \textbf{Value} & \textbf{Source} \\ \hline
lambda\_vac1: TESA SG-L vacuum wavelength, nm \newline Report: OFS-2024-0006 \newline Date: 2024-05-02 & 543.5153892 & \cite{white2025provenance} \\ \hline
lambda\_vac2: TESA SG-O vacuum wavelength, nm \newline Report: OFS-2024-0005 \newline Date: 2024-04-30 & 611.9703724 & \cite{white2025provenance} \\ \hline
lambda\_vac3: Spectra 117A-1 HeNe vacuum wavelength, nm \newline Report: OFS-2024-0002 Date: 2024-05-08 & 632.9909778 & \cite{white2025provenance} \\ \hline
$T_{air}$: air temperature, degC & 20.00 & Absent, value is taken from ISO 1:2002 \\ \hline
$P_{air}$: air pressure, Pa & 101325.0 & Absent, standard atmosphere pressure is taken \\ \hline
h: relative humidity & 0.50 & Absent, neutral value is selected \\ \hline
xCO2: CO2 mole fraction, ppm & 450.0 & Default in the Ciddor (1996) equation \\ \hline
m: integer fringe order (method of exact fractions) & 22123 & Absent, back-calculated from the 7 mm nominal length $ 2*L_{nom}*n / lambda_{vac3}$. Where n is the result of Ciddor equation calculation: 1.000272 \\ \hline
f: fractional fringe order (observed) & 0.23675 & Absent, selected to converge final results \\ \hline
alpha: thermal expansion coefficient of tungsten carbide, $K^{-1}$ & 0.00000423 & \cite{white2025provenance} \\ \hline
$T_{part}$: part temperature, degC & 20.001 & Absent, synthetic value is set to be physically realistic and produce non-zero thermal correction \\ \hline
$T_{ref}$: ISO 1 reference temperature, degC & 20.00 & Reference temperature for length metrology \\ \hline
$L_{nom}$: nominal gauge block length (7 mm), nm & 7000000 & \cite{white2025provenance} \\ \hline
\end{tabular}
\end{table}

As indicated in the Source column of Table \ref{tab:metrology-inputs}, seven of the thirteen inputs required for the calculation were not taken from the original publication \cite{white2025provenance} and were instead reconstructed based on physical constraints, international standards (e.g., ISO 1:2002), or analytical back-calculation. One of these, the fractional fringe order f, was explicitly selected to converge on the reference publication's reported length rather than independently derived or measured; because this parameter was chosen to match a target output, the resulting calculation is circular with respect to that specific value and should be read as a reconstruction and independent re-implementation of the reported procedure under stated assumptions, not as an independent empirical validation of the original publication's result. This distinction does not diminish the CPG demonstration itself: the framework captured full computational lineage, including every assumed and back-calculated input, in a form that makes the conditional nature of the result explicit and auditable rather than hidden inside an opaque final figure. This is, in fact, the central practical argument for microscopic provenance in scientific reconstruction: where an original computation cannot be exactly reproduced because required inputs were never published, the CPG still records precisely which values were sourced, assumed, or back-calculated, so that the reconstructed length in Section~\ref{Sec: results} is transparently conditional on the assumptions in Table~\ref{tab:metrology-inputs} rather than presented as an unqualified reproduction. By capturing the entire computational lineage from initial inputs to final outputs, including all transient state transformations and intermediate results, the CPG maintains the integrity of the computational process and ensures that the origins of every variable remain transparent and verifiable for subsequent peer review. To determine the refractive index of moist air (n) for wavelength correction, modified Edlen \cite{birch1993updated} formula was encapsulated. The computational sequence begins with the calculation of the primary laser wavenumber:
\begin{equation}
    \sigma=\frac{1}{\lambda_{vac}}
\end{equation}

Subsequently, the standard refractivity for dry air is determined ($n_s$):
\begin{equation}
n_s=8342.54+\frac{2406147}{130 - \sigma^2}+\frac{15998}{38.9 - \sigma^2}
\end{equation}

A $CO_2$ correction is applied for $n_s$ to yield $n_{sCO2}$, where $xCO_2$ is the actual $CO_2$ concentration within the interferometer:
\begin{equation}
n_{sCO2}=1+0.534e^{-6}*(xCO_2-450)
\end{equation}

Temperature and pressure corrections ($n_{tp}$) are then applied to $n_{sCO2}$, where P and T denote the pressure and temperature respectively:
\begin{equation}
n_{tp}=n_{sCO2}*P*\frac{(1+P*1e^{-8}*(0.601 - 0.00972*T))}{96095.43*(1+0.003661*T)}
\end{equation}

Water vapor enhancement factor:
\begin{equation}
f_{enh}=1.00070+3.7e^{-8}*P-1.24e^{-7}*T^2
\end{equation}

Saturation vapor pressure is calculated using Wexler formula \cite{wexler1976vapor}, where $T_K$ is temperature in Kelvin and constants: $A=0.000012378847, B=-0.019121316, C=33.93711047, D=-6343.1645$:
\begin{equation}
s_{vp}=e^{(A*T_K^2 + B*T_K + C+\frac{D}{T_K})}
\end{equation}

Calculate water vapor correction to refractivity, where h is humidity:
\begin{equation}
n_{v}=-(3.8020-0.0384*\sigma^2)*h*s_{vp}*f_{enh}+B*T_K+C+\frac{D}{T_K}
\end{equation}

Total refractive index of air:
\begin{equation}
n=1+(n_{tp}+n_v)*1e^{-8}
\end{equation}

Resulting wavelength in air:
\begin{equation}
\lambda_{air}=\frac{\lambda_{vac}}{n}
\end{equation}

Utilizing the resulting wavelength in air, alongside the integer (m = 22123) and fractional (f = 0.23675) fringe orders, the raw interferometric length ($L_{raw}$) was calculated:
\begin{equation}
L_{raw}=\frac{(m+f)*\lambda_{air}}{2}
\end{equation}

A subsequent thermal correction was applied to adjust the raw length to the ISO 1:2002 reference temperature $T_{ref}$, accounting for the specific part temperature $T_{part}$ and the thermal expansion coefficient of the tungsten carbide ($\alpha = 4.23 * 10^{-6} K^{-1}$). The final calibrated length is determined as follows:
\begin{equation}
L_{cal}=\frac{L_{raw}}{1+\alpha(T_{part}-T_{ref})}
\end{equation}

Every intermediate calculation was automatically recorded in the CPG as the environmental inputs were processed. The final framework execution successfully computed a length deviation ($\Delta L$) of +2 nm from the nominal length of 7,000,000 nm. This figure is reported as a point value without an associated measurement uncertainty; a complete metrological treatment following the Guide to the Expression of Uncertainty in Measurement \cite{jcgm_gum} would propagate the uncertainty of each input (the calibrated wavelengths, the environmental readings, and the assumed values in Table~\ref{tab:metrology-inputs}) through to $\Delta L$, and we flag this propagation as outside the scope of the present demonstration rather than as a result already established here. By capturing the entire operational track in a single snapshot, the CPG provides a granular, atomic-level record of the evolution of physical constants and sensor readings through sequential algebraic integration. This capability directly addresses the requirement for machine-readable provenance in metrology, transforming the abstract concept of metrological traceability into a verifiable, serialized computational graph.

\subsection{Hydrological Models Evaluation}

The final evaluation demonstrates the framework’s capacity within the hydrological domain, specifically replicating the metadata management study of the mesoscale Hydrologic Model (mHM) as documented in \cite{villamar2025metadata}. In this context, the model’s performance is evaluated by simulating river discharge for the Moselle River basin upstream of Perl (11,500 $km^2$). While the original publication contains 14 simulations, inconsistencies such as varying temporal ranges and missing output files required a filtering process to identify viable datasets. Data were extracted from the \texttt{b*\_daily\_discharge.out} files, the detailed contents of which are summarized in Table \ref{tab:hydrological-inputs}. While the original simulations involve different date ranges and gauges (\#00398 and \#00333), the evaluation period for this study was standardized to [1991-01-01; 1991-06-30] using gauge \#00398. This specific configuration was chosen to maximize the number of overlapping simulations and ensure a consistent baseline for comparative analysis. In cases where simulation folders contained multiple output files, selection was based on the alignment of date ranges and gauge availability.

This case study aims to quantify model performance and identify the optimal simulation through the implementation of the Kling-Gupta Efficiency (KGE) metric \cite{gupta2009decomposition}, which was encapsulated within the framework’s tracking logic. The KGE is a composite objective function that integrates three mathematically distinct components of the hydrograph: the Pearson correlation coefficient (r), the variability ratio ($\alpha$), and the bias ratio ($\beta$). The framework calculates this metric as the Euclidean distance from the ideal point within the KGE space. The Pearson correlation is determined using the following formula, where $d_{obs}$ and $d_{sim}$ denote, respectively, the observed and simulated discharge values expressed as deviations from their own series mean:
\begin{equation}
r=\frac{\sum({d_{obs}*d_{sim})}}{\sqrt{\sum{d_{obs}^2}*\sum{d_{sim}^2}}}
\end{equation}

The variability ratio ($\alpha$) is determined as follows; because $\sum d_{sim}^2$ and $\sum d_{obs}^2$ are, up to a constant factor, the sample variances of the simulated and observed series, $\alpha$ is numerically equivalent to the ratio of their standard deviations, $\sigma_{sim}/\sigma_{obs}$:
\begin{equation}
\alpha=\sqrt{\frac{\sum{d_{sim}^2}}{\sum{d_{obs}^2}}}
\end{equation}

The bias ratio ($\beta$) is subsequently calculated, where $\mu$ denotes the mean of the observed and simulated values:
\begin{equation}
\beta=\frac{\mu_{sim}}{\mu_{obs}}
\end{equation}

The final KGE is derived using the following objective function:
\begin{equation}
KGE=1-\sqrt{(r-1)^2+(\alpha-1)^2+(\beta-1)^2}
\end{equation}

To assess model performance, the resulting KGE values were evaluated against established estimation criteria. A value of $KGE = 1$ indicates a perfect match between the simulated and observed data, while values exceeding the range of $0.3$ to $0.5$ are typically considered behavioral or good in various hydrological studies. Furthermore, any $KGE > 0$ represents positive predictive skill, indicating the model performs better than the mean of the observations.

In this study, the KGE was calculated for 12 of the 14 simulations \cite{villamar2025metadata}, with the optimal model subsequently selected from this viable subset. While original work served as a benchmark, it omitted the KGE values and the granular tracing of their derivation. To address this lack of transparency, the initial CPG was generated for the primary simulation (Case 0), while the remaining cases were processed using the framework's native support for simulations without source-code intervention. This mechanism enables the substitution of input variables through automated snapshot duplication, ensuring that each variant is executed within a controlled, independent state. Despite the omission of these specific details in the original benchmark paper, this research successfully executed the entire KGE calculation process, capturing the full computational picture including input values, interim calculations, and the final result as a verifiable CPG. Consequently, every computational step from the derivation of means ($\mu$) and deviations ($d$) to the final square-root transformation was encapsulated within the CompProv framework. These subsequent KGE values were computed using the initially created CPG structure, proving its self-containing nature and demonstrating how the framework can retroactively provide reproducibility of calculations. Following the completion of all calculations, a comparative analysis was performed to identify the optimal hydrological model. This end-to-end instrumentation ensures that the model's statistical performance is not merely a reported figure, but a fully traceable and reproducible computational artifact.

\begin{figure}[htbp]
    \centering
    \includegraphics[width=0.8\linewidth]{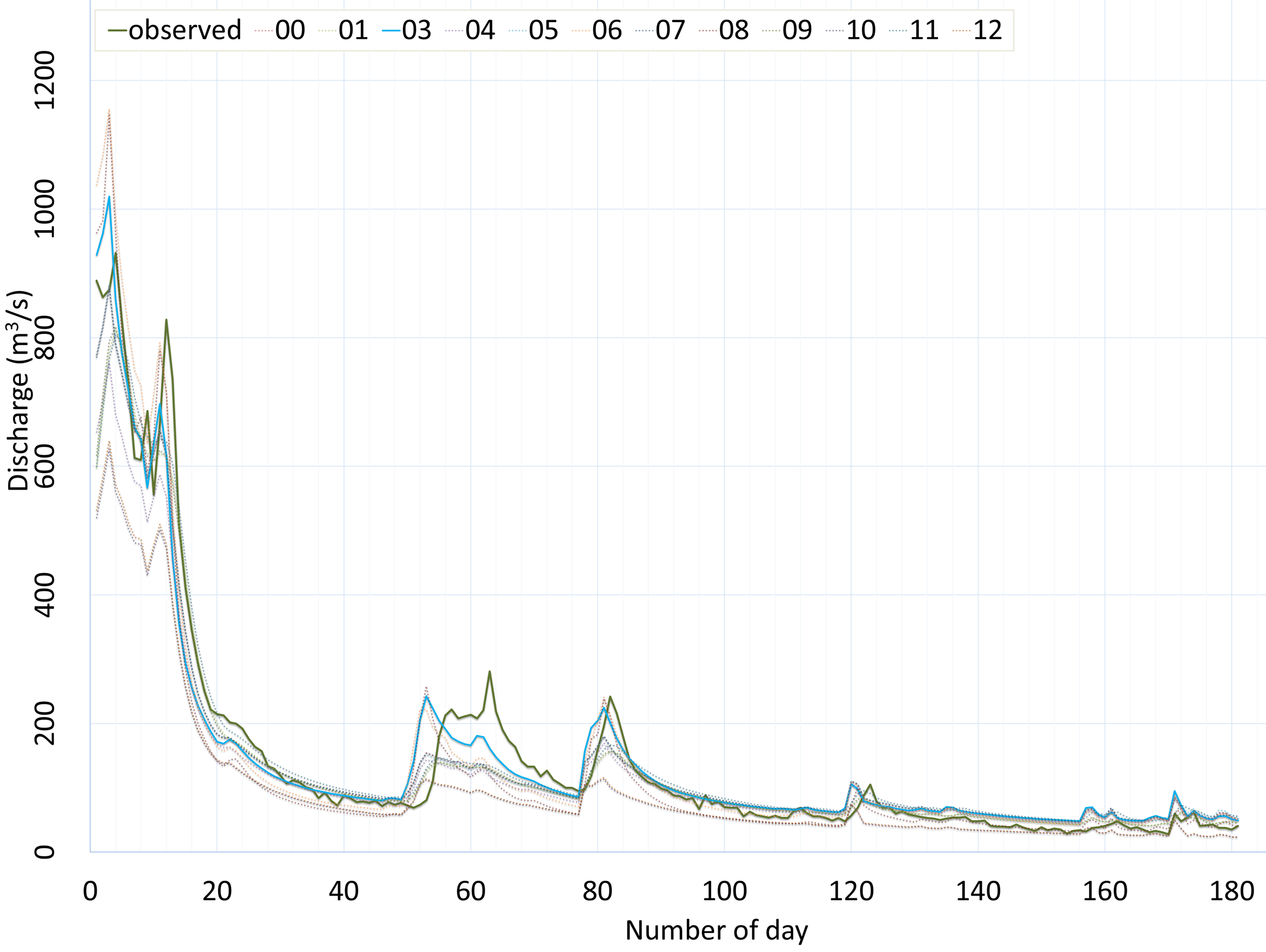}
    \caption{Visualized discharge simulations}
    \label{fig:hydrological-datachart}
\end{figure}

The calculated KGE values, along with the corresponding input details, are presented in Table \ref{tab:hydrological-inputs}. The simulation for Case 03 demonstrated the highest performance, yielding a KGE of 0.93911746. A comparative visualization of all simulations is provided in Figure \ref{fig:hydrological-datachart}. In this chart, the observed values are indicated by a bold green line, while the Case 03 results are represented by a solid blue line. To maintain visual clarity and prevent over-plotting, the remaining simulation results are displayed as dashed lines. The X-axis reflects the number of days since 1991-01-01, and the Y-axis displays discharge in $m^3/s$.

\begin{table}[htb]
\centering
\caption{River discharge simulations details and results}
\label{tab:hydrological-inputs}
\begin{tabular}{|c|c|c|c|c|c|} 
\hline
\textbf{Case} & \textbf{Date from} & \textbf{Date to} & \textbf{Information} & \textbf{Evaluated}  & \textbf{KGE} \\ \hline
00 & 1990-07-01 & 1991-06-30 & b1\_daily\_discharge.out & + &  0.86868599 \\ \hline
01 & 1991-01-01 & 1991-12-31 & b1\_daily\_discharge.out & + &  0.85695795 \\ \hline
02 &            &            & Absent b*\_daily\_discharge.out & - &   \\ \hline
03 & 1990-07-01 & 1991-06-30 & b1\_daily\_discharge.out & + & \textbf{0.93911746}  \\ \hline
04 & 1991-01-01 & 1992-12-31 & b2\_daily\_discharge.out & + &  0.72514941 \\ \hline
05 & 1990-07-01 & 1991-06-30 & b3\_daily\_discharge.out & + & 0.87020740 \\ \hline
06 & 1990-07-01 & 1991-06-30 & b1\_daily\_discharge.out & + & 0.90000436 \\ \hline
07 & 1991-01-01 & 1991-12-31 & b1\_daily\_discharge.out & + & 0.86509116 \\ \hline
08 & 1990-07-01 & 1991-06-30 & b1\_daily\_discharge.out & + & 0.87371946 \\ \hline
09 & 1990-07-01 & 1991-06-30 & b1\_daily\_discharge.out & + & 0.85338966 \\ \hline
10 & 1990-07-01 & 1991-06-30 & b1\_daily\_discharge.out & + & 0.51069941 \\ \hline
11 & 1990-01-01 & 1993-12-31 & b6\_daily\_discharge.out & + & 0.88107321 \\ \hline
12 & 1990-07-01 & 1991-06-30 & b1\_daily\_discharge.out & + & 0.52664834 \\ \hline
13 &            &            & Absent b*\_daily\_discharge.out & - & \\ \hline
\end{tabular}
\end{table}

The resulting CPG captured the entire statistical transformation of the discharge time-series, preserving the functional continuity between raw observations and final metrics. By maintaining an atomic record of the sums of squares and cross-products, the framework ensures that the performance of a hydrological model is no longer an opaque scalar, but a fully auditable result tied directly to the original data. By successfully reproducing and documenting the calculations omitted in the benchmark study, this research provides a transparent mechanism for verifying numerical integrity. Consequently, the reliance on pre-computed, black-box statistics is replaced by a verifiable computational lineage, fulfilling the core requirements for reproducible science in complex environmental modeling.

\subsection{Complexity Analysis}\label{Subsec: complexity}
For acyclic algorithms, the expected theoretical complexity of CPG calculation is identical to the baseline execution's inherent time and memory complexity. If the underlying algorithm features a time complexity of $T(N)$, the corresponding CPG construction scales at $O(T(N))$.

However, for cyclic or iterative algorithms, a naive step-by-step recording causes memory complexity to grow linearly with the iteration count. Even if the baseline algorithm operates in constant memory ($O(1)$), the continuous accumulation of execution nodes forces the graph's space complexity into $O(N)$. To empirically validate these dynamics, a Monte Carlo $\pi$ calculation algorithm was implemented (Algorithm~\ref{alg:pi-calculation}) and evaluated across a broad range of steps ($N$). The time complexity for the algorithm is $O_T(N)$. The algorithm was executed for $N \in [10 000;250 000]$ with a step of $1000$. To mitigate random fluctuations related to internal Java processes like Garbage Collector (GC), the initial heap size was set to 8GB and GC was manually triggered before every run. The execution was done using personal computer (PC). PC hardware parameters are omitted because only the relative dynamics are relevant.

\begin{algorithm}[ht]
\caption{$\pi$ calculation algorithm}
\label{alg:pi-calculation}

\KwIn{A total number of points $N$ (totalPoints)}
\KwOut{Estimated value of $\pi$}

\BlankLine
$counter \leftarrow \text{0}$\;
\For{$i \leftarrow 0$ \KwTo $N$}{
    $x \leftarrow \text{random(0; 1)}$\;
    $y \leftarrow \text{random(0; 1)}$\;
    
    \BlankLine
    $distSquared \leftarrow x^2 + y^2$\;
    $inCircle \leftarrow 1 - round(distSquared, DOWN)$\;
    $counter \leftarrow counter + inCircle$\;
}

\BlankLine
$ratio \leftarrow counter / N$\;
$\pi_{estimated} \leftarrow ratio * 4$\;

\Return $\pi_{estimated}$\;
\end{algorithm}

\begin{figure}[htbp]
    \centering
    \includegraphics[width=\linewidth]{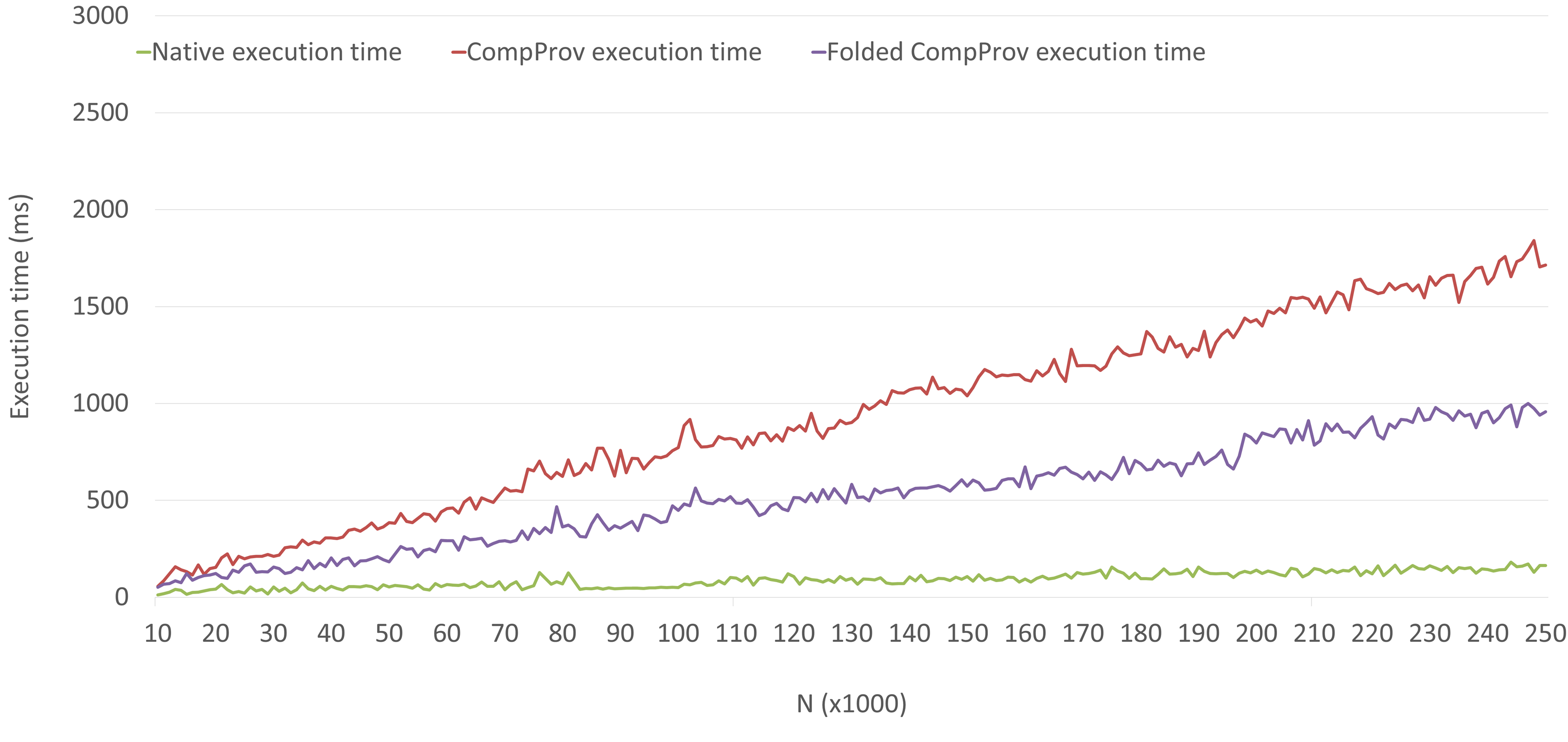}
    \caption{Processing time}
    \label{fig:processing-time-chart}
\end{figure}

As illustrated in Figure~\ref{fig:processing-time-chart}, both the native execution (green line) and the framework-wrapped tracking (red line) exhibit a strict linear relationship between execution time and the scale of $N$. However, the framework-wrapped timeline features a significantly steeper slope compared to the near-flat baseline trajectory. This divergence in their slopes highlights the cumulative nature of the tracking cost, where a fixed per-iteration computational overhead is introduced to generate and manage graph elements as the iteration scale grows.

\begin{figure}[htbp]
    \centering
    \includegraphics[width=0.9\linewidth]{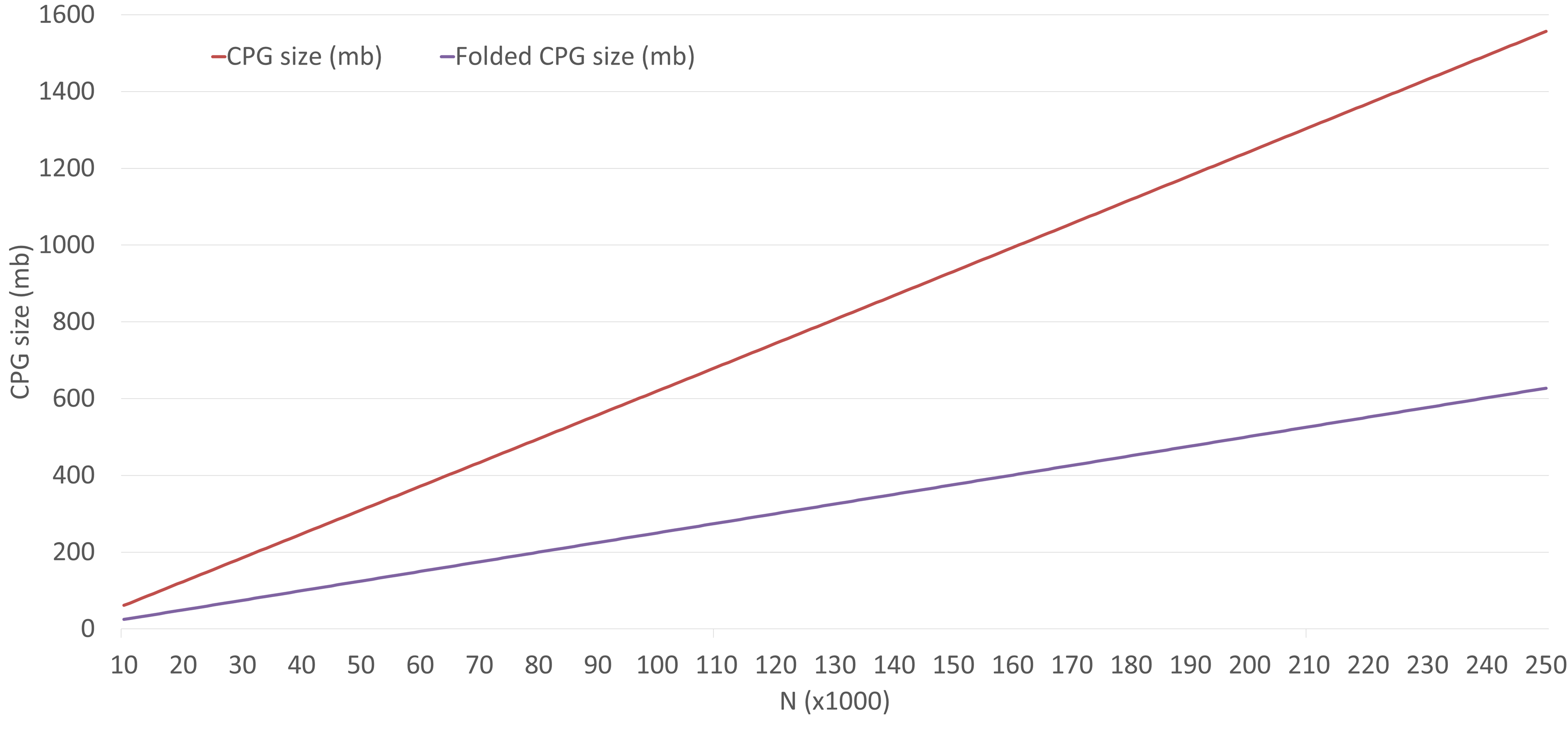}
    \caption{CPG size dynamics}
    \label{fig:cpg-size}
\end{figure}

Figure~\ref{fig:cpg-size} highlights the memory amplification inherent to cyclic graph expansion. The algorithm's own working-variable memory footprint is $O_M(1)$, since the loop in Algorithm~\ref{alg:pi-calculation} reuses the same few scalar variables on every iteration; the CPG artifact that CompProv produces while recording that same execution, in contrast, exhibits $O_M(N)$ growth, as shown by the red line in Figure~\ref{fig:cpg-size}. This is not a contradiction: the two quantities describe different things, the algorithm's transient working memory and the size of the persistent provenance record it generates. The growth is dictated by the acyclic nature of the graph, as every iteration is expanded into new, immutable nodes that reflect the state of variables at that point in time rather than overwriting a single running value.

To mitigate memory consumption trends and reduce execution overhead, a subgraph folding technique is introduced. The framework records the complete, fully granular computation, every intermediate operation and value, for exactly one representative pass through the repeated loop body; this fully recorded pass is retained as a reusable Template. For every subsequent iteration, the same Template logic is invoked again with that iteration's own input, but the CPG records only the input and the resulting output for that iteration, not its intermediate values, since those are already structurally identical to the Template and would otherwise be duplicated N times at no informational gain.

Comparative benchmarks validate that this abstraction yields substantial efficiency gains, as evidenced by the performance trajectories of the folded execution context (purple line in Figure~\ref{fig:processing-time-chart} and Figure~\ref{fig:cpg-size}). By condensing multiple internal operations per iteration into a single template invocation, the framework dramatically minimizes the exported serialization snapshot size. Because intermediate internal operations are bypassed and no longer preserved in memory, the cumulative time overhead is drastically reduced as well, preventing the sharp slope divergence observed in the fully unfolded tracking. Detailed performance analysis is provided in Table~\ref{tab:pi_benchmark}. 

\begin{table}[htbp]
\centering
\caption{Comprehensive Normalized Performance, Tracing Overhead, and Memory Analysis ($\pi$ Calculation Benchmark)}
\label{tab:pi_benchmark}
\begin{tabular}{p{4cm}ccc}
\toprule
\textbf{Metric} & \textbf{Native (JPC)} & \textbf{CompProv (Unfolded)} & \textbf{CompProv (Folded)} \\
\midrule
\textbf{Execution Time per 1k Ops (ms)} & & & \\
~~Mean $\pm$ SD & $0.83 \pm 0.41$ & $7.46 \pm 0.69$ & $4.24 \pm 0.65$ \\
~~Median & $0.69$ & $7.34$ & $4.04$ \\
\midrule
\textbf{Absolute Overhead per 1k Ops (ms)} & & & \\
~~Mean $\pm$ SD & — & $6.63 \pm 0.56$ & $3.41 \pm 0.52$ \\
~~Median & — & $6.56$ & $3.35$ \\
\midrule
\textbf{Relative Overhead (Multiplier)} & & & \\
~~Mean $\pm$ SD & — & $10.25\times \pm 3.01\times$ & $5.75\times \pm 1.56\times$ \\
~~Median & — & $10.44\times$ & $5.84\times$ \\
~~Min / Max Range & — & $3.11\times$ -- $17.97\times$ & $1.57\times$ -- $9.29\times$ \\
\midrule
\textbf{CPG Memory Footprint per 1k Ops} & & & \\
~~Mean $\pm$ SD & — & $6352.94 \pm 16.76$ KB & $2562.60 \pm 6.81$ KB \\
~~(Equivalent in MB) & — & $\approx 6.20$ MB & $\approx 2.50$ MB \\
\bottomrule
\end{tabular}
\end{table}

Importantly, optimizing the global graph via folding does not sacrifice structural inspectability. Because every folded iteration was itself computed by invoking the same Template logic, an auditor who needs to inspect a specific suspicious iteration can set the Template's input to that iteration's recorded input value and re-execute it; every intermediate value that the folded record omitted becomes visible again in this re-execution, and the recomputed output can be checked against the value the CPG originally recorded for that iteration as a consistency check. This is a targeted, deterministic re-execution of that one iteration through the preserved Template, not a replay of an internal trace that was never stored, and it therefore presumes the Template is itself deterministic and free of external side effects.

However, while the $\pi$ stress benchmark provides valuable insights into the baseline behavior of the framework, it represents a workload dominated by a massive number of lightweight iterations with low payload complexity. In such scenarios, the relative tracing overhead inevitably appears inflated because the native execution time of each individual operation is close to negligible. To confirm that this overhead becomes less pronounced as the complexity of the underlying computation increases, a second $\pi$ estimation stress-test was implemented. This approach computes the definite integral of:
\begin{equation}
f(x)=\frac{1}{\sqrt{1-x^2}}
\end{equation}
over the interval $[0, 1]$. Because $f(x)$ becomes unbounded as $x$ approaches 1, traditional rejection sampling against a finite bounding box is inapplicable. Instead, the framework evaluates the integration using the average-value method:
\begin{equation}
\int_{a}^{b} f(x)=\frac{(b-a)}{N}\sum_{i=1}^{N} f(x_i)
\end{equation}

for $x_i$ drawn uniformly from $[0, 1)$. While the singularity at $x=1$ is integrable (yielding the exact value of $arcsin(1)-arcsin(0)=\pi/2$ , the variance is infinite because $f^2$ is non-integrable near the boundary. As a result, this single-loop implementation is significantly more complex and statistically less effective, exhibiting slower and noisier convergence than standard polynomial estimators. However, the primary aim of this benchmark was to demonstrate how increased loop complexity affects overhead and memory consumption under different execution models. Unlike the $\pi$ stress test, the $\arcsin$ algorithm features a higher operational complexity within its payload, executing more intensive mathematical computations during each step. A detailed algorithm comparison is provided in Table~\ref{tab:algorithm_comparison}. 
\begin{table}[htbp]
\centering
\caption{Structural and Performance Comparison of Benchmark Algorithms}
\label{tab:algorithm_comparison}
\begin{tabular}{lcc}
\toprule
\textbf{Characteristic / Feature} & \textbf{$\pi$ Stress Benchmark} & \textbf{$\arcsin$ Benchmark} \\
\midrule
Number of Inputs & 2 & 1 \\
Number of Variables & 6 & 4 \\
Number of Operations per Loop Iteration & 5 & 4 \\
\midrule
\textbf{Native Execution Time per 1k Ops (ms)} & & \\
~~Mean $\pm$ SD & $0.83 \pm 0.41$ & $3.00 \pm 0.20$ \\
~~Median & $0.69$ & $2.97$ \\
\bottomrule
\end{tabular}
\end{table}

\begin{figure}[htbp]
    \centering
    \includegraphics[width=\linewidth]{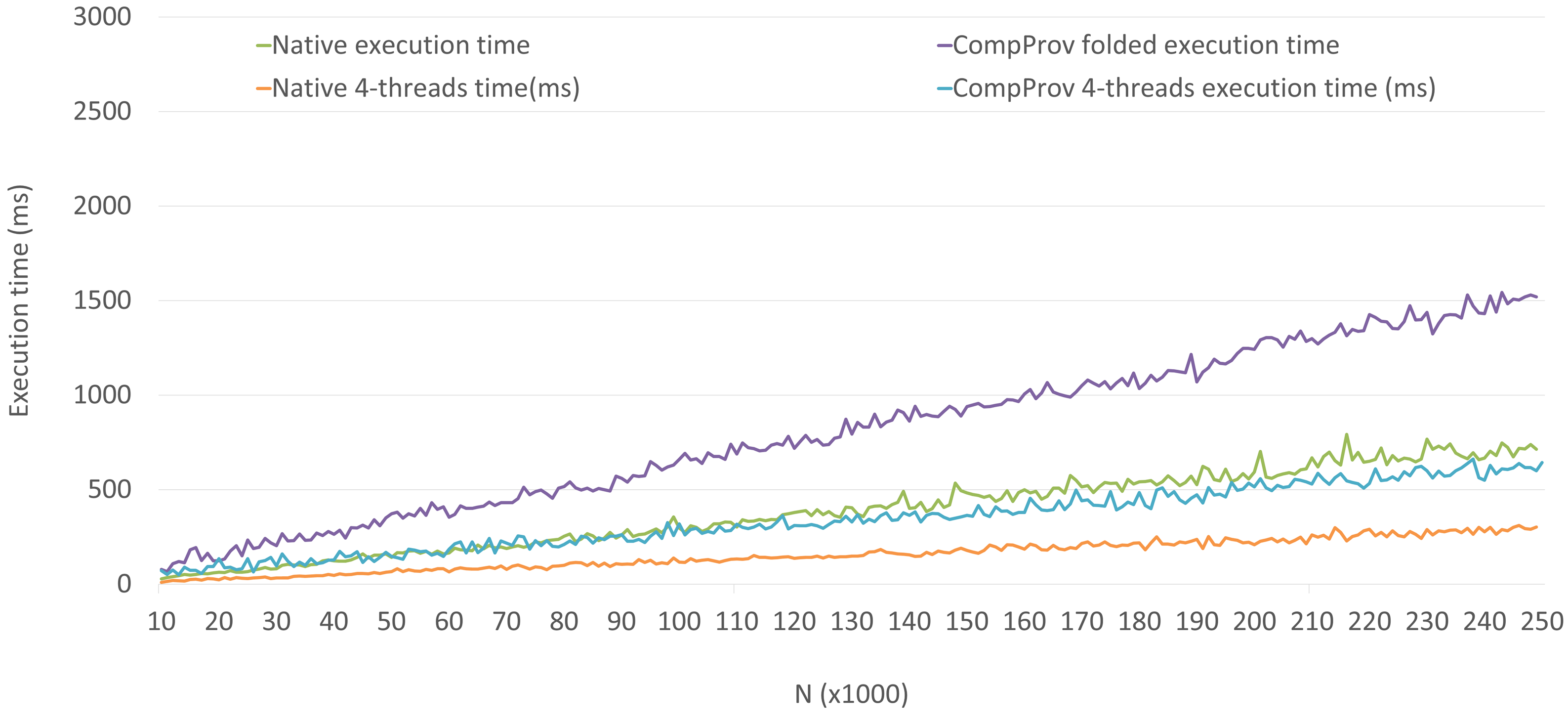}
    \caption{Processing time}
    \label{fig:processing-time-chart2}
\end{figure}

\begin{figure}[htbp]
    \centering
    \includegraphics[width=\linewidth]{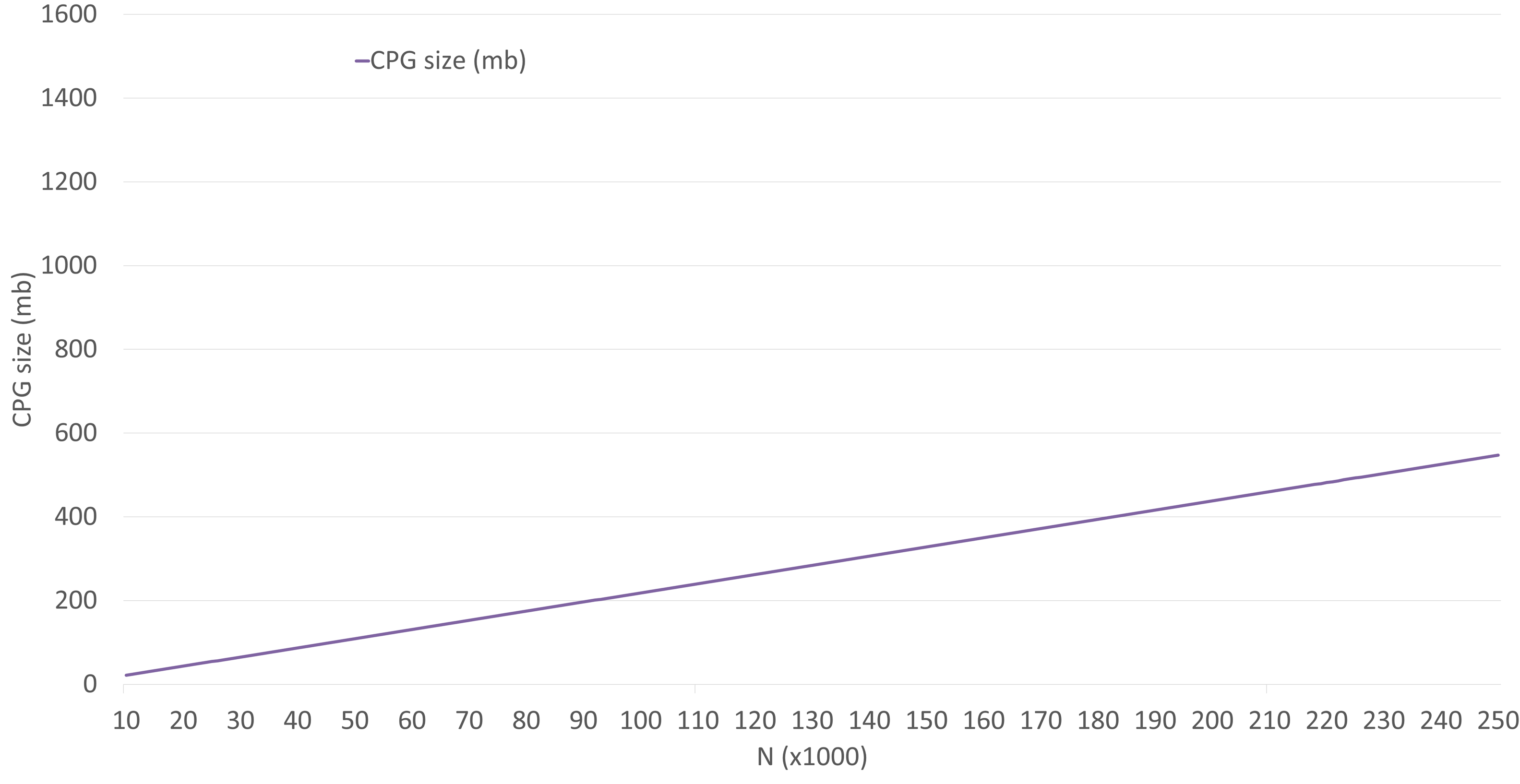}
    \caption{CPG size dynamics}
    \label{fig:cpg-size2}
\end{figure}

Comparative benchmarks for the computationally intensive $\arcsin$ workload are presented in Figures~\ref{fig:processing-time-chart2} and~\ref{fig:cpg-size2}, where the green and purple trajectories represent the normalized single-threaded native execution and the folded framework-wrapped mode, respectively. Due to the higher payload profile of the algorithm, the gap between the native and tracing performance curves narrows significantly, reflecting a substantial amortization of the relative tracking overhead. Concurrently, the growth rate of the CPG memory footprint per 1,000 operations exhibits near-perfect stabilization, introducing minimal memory inflation. Furthermore, to assess how well the framework's performance scales under concurrency, a 4-threaded execution scenario was evaluated. The results, illustrated via the orange trajectory for multi-threaded native execution and the blue trajectory for the 4-threaded folded CompProv execution, demonstrate parallel scaling and throughput comparable to native execution under this workload.

Quantitative analysis of the experimental data compiled in Tables~\ref{tab:pi_benchmark} and~\ref{tab:arcsin_benchmark} further confirms these observations. Specifically, the median relative time overhead drops substantially from $5.84\times$ in the $\pi$ benchmark to just $2.13\times$ for the $\arcsin$ workload. This reduction is primarily driven by the computational profile of the payload: as the native execution time per 1,000 operations scales from a median of $0.69$~ms to $2.97$~ms, the median absolute tracing overhead remains practically invariant, shifting negligibly from $3.35$~ms to $3.30$~ms. This decoupled behavior indicates that heavier internal operations do not introduce additional absolute tracing penalties, thereby naturally compressing the relative overhead. Concurrently, the normalized CPG memory footprint is reduced, an outcome directly dictated by the structural simplification of the input data vector, as the $\pi$ algorithm requires two random boundary inputs whereas the $\arcsin$ loop relies on a single input. Finally, the parallel efficiency evaluation reveals nearly identical multi-threaded speedup factors for both the native execution ($2.47\times$) and the CompProv-wrapped environment ($2.39\times$). This minimal divergence under concurrent scaling indicates that the wrapping overhead does not meaningfully erode parallel throughput for this workload; it demonstrates performance scalability under this specific concurrency pattern rather than a formal verification of thread safety or reentrancy, consistent with the scope stated in Section~\ref{Sec: methods}.

It is important to emphasize that the absolute CPG memory footprint directly correlates with the length of the identifier strings and metadata descriptors assigned to the graph components. To ensure that our evaluation reflects production-grade conditions rather than synthetic best-case scenarios, we deliberately utilized highly descriptive, human-readable, and realistic variable names throughout both benchmarks. Consequently, the reported memory metrics represent a conservative upper bound, whereas production environments employing compacted or minified identifiers could achieve even lower memory overhead.

\begin{table}[htbp]
\centering
\caption{Comprehensive Multi-Threading Normalized Performance, Overhead, and Memory Analysis ($\arcsin$ Benchmark)}
\label{tab:arcsin_benchmark}
\footnotesize
\begin{tabular}{p{3cm}ccc}
\toprule
\textbf{Metric} & \textbf{Single-Thread (1-T)} & \textbf{Multi-Thread (4-T)} & \textbf{Parallel Speedup} \\
\midrule
\textbf{Native Execution Time per 1k Ops (JPC)} & & & \\
~~Mean $\pm$ SD (ms) & $3.00 \pm 0.20$ & $1.22 \pm 0.12$ & $\mathbf{2.48\times \pm 0.26\times}$ \\
~~Median (ms) & $2.97$ & $1.20$ & $\mathbf{2.47\times}$ \\
\midrule
\textbf{CompProv Folded Time per 1k Ops} & & & \\
~~Mean $\pm$ SD (ms) & $6.47 \pm 0.80$ & $2.87 \pm 0.75$ & $\mathbf{2.32\times \pm 0.32\times}$ \\
~~Median (ms) & $6.25$ & $2.60$ & $\mathbf{2.39\times}$ \\
\midrule
\textbf{Absolute Tracing Overhead per 1k Ops (ms)} & & & \\
~~Mean $\pm$ SD & $3.47 \pm 0.78$ & $1.65 \pm 0.73$ & — \\
~~Median & $3.30$ & $1.41$ & — \\
\midrule
\textbf{Relative Tracing Overhead (Multiplier)} & & & \\
~~Mean $\pm$ SD & $2.16\times \pm 0.27\times$ & $2.37\times \pm 0.64\times$ & — \\
~~Median & $2.13\times$ & $2.19\times$ & — \\
~~Min / Max Range & $1.66\times$ -- $3.76\times$ & $1.73\times$ -- $7.60\times$ & — \\
\midrule
\textbf{CPG Memory Footprint per 1k Ops} & & & \\
~~Mean $\pm$ SD & $2237.23 \pm 5.69$ KB & — & — \\
~~(Equivalent in MB) & $\approx 2.18$ MB & — & — \\
\bottomrule
\end{tabular}
\end{table}

\section{Discussion}\label{Sec: discussion}

\begin{figure}[htbp]
    \centering
    \includegraphics[width=0.8\linewidth]{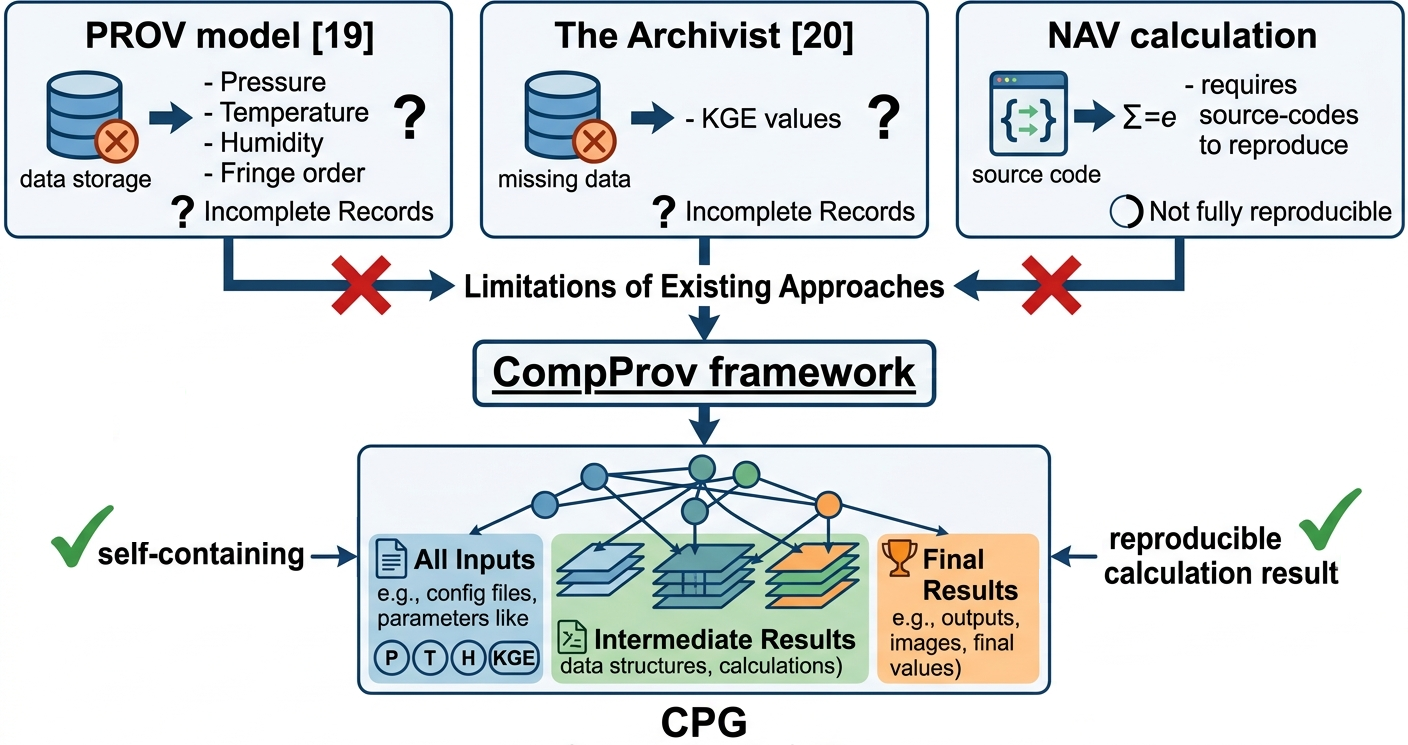}
    \caption{CompProv benchmarks}
    \label{fig:benchmarks}
\end{figure}

The three case studies in this work were not designed to push CompProv to its 
limits. They were designed to answer a more fundamental question: can a framework 
that wraps numbers at the arithmetic level produce a provenance record complete 
enough to replay, audit, and substitute inputs across domains that share no 
structural similarity? The results confirm that it can, and the reasons why that 
matters extend well beyond the three workflows tested here 
(Figure~\ref{fig:benchmarks}).

The reproducibility crisis in computational science~\cite{baker20161, 
stupple2019reproducibility} is typically described as a problem of missing 
code, missing data, or missing documentation. The gauge block calibration 
result points to a fourth category: missing arithmetic. The original published 
procedure~\cite{white2025provenance} was complete enough to describe the 
method, but seven of the thirteen numerical inputs required to execute that method were never 
recorded: air temperature, air pressure, relative humidity, the CO2 mole fraction, 
the integer and fractional fringe orders, and the part temperature. No workflow-level provenance system could have surfaced this gap, because 
those systems operate on files and processes, not on the values that flow through 
them. CompProv exposed the gap because its architecture requires every variable 
entering a computation to be explicitly wrapped and tagged before any operation 
can proceed. A missing input is not an oversight that can be reconstructed 
after the fact, it is a structural absence that the CPG makes visible and 
documents explicitly, as Table~\ref{tab:metrology-inputs} records for each of the seven affected variables. The hydrological evaluation revealed the same problem 
from a different direction. The original benchmark~\cite{villamar2025metadata} 
provided raw simulation output but reported no KGE values and gave no 
description of how they should be derived from the data. Reproducing those 
values required manually reverse-engineering the file structure before a single 
formula could be applied. Once CompProv was applied to Case 00 and the initial 
CPG was generated, the remaining 11 evaluated simulations were processed through snapshot 
duplication without writing new computational logic. The best-performing 
simulation, Case 03, yielded a KGE of 0.939, a result that now exists as a 
fully traceable artifact tied directly to the raw discharge observations, rather 
than a number in a table with no derivation attached. These two cases together 
make a concrete argument: the problem is not always that researchers fail to 
share their code or data. Sometimes the intermediate numerical record was simply 
never captured, and no amount of documentation can recover what was never stored.

Table~\ref{tab:comparison} in Section~\ref{Sec: related} positioned CompProv 
against coarse-grained lineage systems, script-level provenance tools, 
automatic-differentiation engines, and numerical-accuracy diagnostics. Of 
these, tape-based automatic differentiation remains the closest 
architectural relative, since it also builds fine-grained computation graphs 
through operator overloading at the arithmetic level, and the comparison is 
also the starkest: an AD tape from a 2024 financial calculation cannot 
answer a 2031 audit question, because the tape no longer exists once the 
gradient it was built to compute has been returned, while a CompProv CPG, 
being a persisted artifact rather than a discarded by-product, can. The 
gauge block and hydrological cases in Section~\ref{Sec: results} give this 
distinction concrete weight: in both, the gap that CompProv exposed, 
undocumented inputs in one, undocumented derivations in the other, is 
precisely the kind of internal detail that no file-, pipeline-, or 
environment-level system in Table~\ref{tab:comparison} was designed to see.

The practical implications of this position in the landscape are clearest in 
the NAV case. Financial institutions performing historical audits face a genuine 
technical problem: the software that produced a valuation several years ago may 
depend on library versions, runtime environments, or external APIs that no longer 
exist in the same form. Reconstructing that environment is expensive and often 
incomplete. A CPG produced at execution time is a self-contained, 
environment-independent record of the operational logic at that moment. It can 
be shared with a regulator to verify mathematical correctness without disclosing 
the proprietary source code that generated it, a distinction that has direct 
commercial and legal relevance. The NAV calculation confirmed a final portfolio 
value of 431,749.17 USD with exact replay fidelity in a fresh environment, and 
the sensitivity analysis demonstrated that input substitution propagates correctly 
through the full graph without any modification to the underlying computational 
logic. For an auditor, this means that hypothetical scenarios, what would the 
valuation have been under different exchange rates, can be explored on a 
historical CPG without access to the original trading system.

That said, the results should be read with an honest account of what the 
framework currently cannot do. All three case studies involve small, 
deterministic, acyclic workflows. The complexity analysis in 
Section~\ref{Subsec: complexity} showed that the $\pi$-estimation algorithm's own working-variable footprint is $O_M(1)$, while the CPG artifact it produces necessarily grows as $O_M(N)$, 
because every iteration creates new immutable graph nodes rather than overwriting existing state. At 250,000 iterations the folded CPG 
reached approximately 550 MB, and a $2.16\times \pm 0.27\times$ median execution-time overhead was measured on the Monte Carlo $arcsin$-integration benchmark even after folding. The 4-threaded benchmark shows wrapped execution scaling in line with native execution under that concurrency pattern, but this linear growth rate and the remaining relative overhead still make the unfolded framework impractical for simulations running millions of iterations, such as those standard in climate modeling, molecular dynamics, or computational fluid dynamics. The replay mechanism also carries a conditional dependency: while the CPG can be replayed without access to the original business logic or proprietary algorithms, it does require both the CompProv wrapper classes and any user-defined domain classes associated with the recorded variables to be present in the replay environment to facilitate successful state deserialization and operational reproducibility. What is truly independent is the domain computation, not the provenance infrastructure itself. Finally, the current Java implementation places a practical barrier for scientific communities that work primarily in Python, and the framework has no mechanism for non-deterministic operations where external values may have changed between the original execution and the replay. These are not reasons to dismiss the approach, they are 
the boundaries of what has been demonstrated, and they define the roadmap for what comes next.

\section{Threats to Validity}\label{Sec: threats}
Several factors may influence the generalizability and accuracy of our findings. We categorize these threats and describe the mitigation strategies employed.

\subsection{External Validity}
The primary threat to external validity is the representativeness of the evaluated algorithms. While our results confirm the theoretical complexity for acyclic processes, highly complex simulations may produce CPGs that strain graph parsing and data ingestion. To mitigate this, we propose partitioning simulations into discrete stages where each step is treated as an independent CPG. A second threat concerns the current implementation's dependence on a single language. The abstract functional mapping that a CPG encodes, Equation~\ref{Eq: functional-mapping}, is language-agnostic in principle, since it describes a relation between input and output identifiers rather than any Java-specific construct. However, the concrete mechanisms that make replay possible, the wrapper classes, the Environment registry, the value-class identifiers, and the operation-descriptor lookup table, are specific to the current Java implementation and have not been reimplemented or evaluated in another language. The claim that CompProv can be ported to C++ or Python without loss of structural integrity is therefore an architectural expectation rather than a demonstrated result, and we qualify it accordingly until a cross-language implementation or replay has been carried out.

\subsection{Internal Validity}
A primary internal threat to validity lies in the computational and memory overhead inherent in microscopic provenance tracking. In our initial evaluations of fully unfolded tracing, we observed a peak median increase of $10.44\times$ in execution time for atomic operations (Table~\ref{tab:pi_benchmark}), which could severely limit the framework's applicability in performance-critical or High-Performance Computing (HPC) environments. 

To mitigate this performance penalty, we successfully designed and integrated a subgraph folding technique. As demonstrated by our benchmark results, this approach successfully amortizes the tracing cost, compressing the median relative overhead down to $5.84\times$ for lightweight tasks and further reducing it to $2.13\times$ for computationally intensive payloads (Table~\ref{tab:arcsin_benchmark}).

Folding itself introduces a secondary internal validity consideration regarding provenance granularity: by grouping iterative steps into unified macro-operations, the framework does not retain the intermediate local variables inside a folded pass, only the input and output of each iteration and the one fully recorded Template pass. To mitigate this loss of granularity, the framework re-executes the preserved Template with a specific iteration's recorded input on demand, recovering its full internal steps as an independent subgraph. This drill-down capability is a targeted re-execution of that specific iteration through the Template rather than a replay of an already-recorded trace, so its correctness depends on the Template being deterministic and free of external side effects, a condition satisfied by the benchmarks in this study but not verified in general.

\subsection{Construct Validity}
Construct validity concerns whether our metrics accurately reflect the framework's utility. Although the analysis of massive CPGs can be challenging, the machine-readable nature of the CPG allows for the development of automated debugging and analysis tools. Future work will focus on CPG fragmentation and advanced analytics engines to ensure that the "transparency-to-overhead" ratio remains favorable for large-scale scientific applications.

\section{Conclusion}\label{Sec: conclusion}
This study addressed the absence of operation-level traceability in existing provenance systems, which record lineage at the level of files, datasets, or pipeline stages while leaving an algorithm's internal arithmetic unrecorded. The proposed CompProv framework closed this gap by encapsulating numerical values in immutable, metadata-carrying wrapper objects that captured every algebraic operation as it executed, producing a serializable CPG. Across three heterogeneous case studies (finance, metrology, and hydrology), deterministic replay reproduced each computed result exactly in a fresh environment, and CPG-based input substitution supported sensitivity analysis without exposing the underlying source code. The complexity analysis further showed that a CPG's memory footprint grows linearly with the number of recorded operations even though the originating algorithm's own working memory remains constant, and that wrapping numerical values incurred up to a $2.16\times \pm 0.27\times$ execution-time overhead even after folding-based mitigation. These findings indicate that coupling metadata capture intrinsically to arithmetic execution is a technically feasible route to auditable-by-design computation across domains with no structural similarity.

The framework's practical value lies in decoupling numerical audit from the disclosure of proprietary business logic and from the survival of a particular software environment, a property relevant to financial auditing, regulated physical measurement, and reproducible environmental modeling. The present evaluation is limited to small, deterministic, acyclic workloads, and the associated overhead currently constrains high-performance computing applications. Future work should pursue CPG fragmentation and compression strategies for large-scale workloads, empirical overhead measurements on domain-specific computations beyond the benchmark used here, and a cross-language implementation to test the portability of the underlying functional mapping. By making the atomic arithmetic of a computation, not merely its inputs and outputs, into a persistent and shareable record, this work provides a concrete foundation for treating numerical integrity as a verifiable property of computation itself.





\section*{Declarations}
\begin{itemize}
    \item \textbf{Funding:} This work was supported by the United Arab Emirates University (UAEU) through the UAEU Program for Advanced Research (UPAR) under Grant 12N174.
    \item \textbf{Competing Interests:} The authors declare no competing interests.
    \item \textbf{Ethics approval and consent to participate:} Not applicable.
    \item \textbf{Consent for publication:} Not applicable.
    \item \textbf{Data availability:} All data generated or analysed during this study are included in this published article and its supplementary information files. The pi-estimation, arcsin-integration, and sensitivity-analysis benchmark data are provided as supplementary spreadsheets (\texttt{pi.xlsx}, \texttt{arcsin.xlsx}, \texttt{sensitivity.xlsx}). The hydrological and metrological case studies are independent reconstructions of scenarios reported in \cite{villamar2025metadata} and \cite{white2025provenance}, respectively; the specific inputs used for these reconstructions are listed in Tables~\ref{tab:hydrological-inputs} and \ref{tab:metrology-inputs} and in the accompanying supplementary spreadsheet (\texttt{hydrology.xlsx}).
    \item \textbf{Materials availability:} Not applicable.
    \item \textbf{Code availability:} The CompProv framework code is available in the repository at \url{https://github.com/compprov/compprov-core/releases/tag/v0.4.2}. The rendering engine is available at \url{https://github.com/compprov/compprov-render/releases/tag/v0.2.0}.
    \item \textbf{Author Contributions:} M.A. conceived the core idea of the study, developed the CompProv framework, implemented the methodology, conducted the experiments, and wrote the original manuscript draft. M.A.A. contributed to the conceptual development of the work, supervised the research, and participated in drafting, reviewing, and editing the manuscript. N.S. provided supervision, contributed to the validation and interpretation of results, and reviewed and edited the manuscript. All authors reviewed and approved the final version of the manuscript.
\end{itemize}

\bibliography{Ref}

\end{document}